\documentclass[amsmath,12pt,amssymb,preprint,prd,aps,nofootinbib]{revtex4}
\usepackage{graphicx} % Required for inserting images
\usepackage{ulem}
\usepackage{color}
\begin{document}
\title{Diffusion and shear viscosity coefficients
of hot isospin asymmetric strange hadronic matter
using a chiral SU(3) model}
\author{Amruta Mishra}
\email{amruta@physics.iitd.ac.in}
\affiliation{Department of Physics, Indian Institute of Technology Delhi, 
Hauz Khas, New Delhi -- 110016, India}
\author{Shujun Zhao}
\email{zhaosj@sophia.ac.jp}
\affiliation{Department of Physics, Sophia University,
Tokyo, 102-8554, Japan}
\author{Tetsufumi Hirano}
\email{hirano@sophia.ac.jp}
\affiliation{Department of Physics, Sophia University,
Tokyo, 102-8554, Japan}
%\vskip -1.5in
%\author{Amruta Mishra}
%\date{June 2026}

\begin{abstract}
We study the diffusion and shear viscosity coefficients of hot isospin
asymmetric strange hadronic matter. The effects due to the baryon density, 
isospin, strangeness, and temperature on the nucleons
and hyperons are studied within a chiral SU(3) model.
The medium modifications of the baryons 
arise due to interactions with the 
mean scalar and vector fields within the model. The thermodynamic
and transport properties are studied in the hot strange hadronic matter.
The diffusion matrix associated with the multiple charges (baryon number, 
isospin, and strangeness), as well as the coefficient
of shear viscosity, are
computed from the Boltzmann equation using first-order
Chapman--Enskog expansion within the relaxation time approximation. 
The diffusion coefficient, $\kappa_{QQ}$ corresponding
to the conserved electric charge has also been studied in the
present work.
There are observed to be significant effects from the isospin asymmetry as well as
the strangeness of the medium on the diffusion coefficients. 
The coefficient of shear viscosity, $\eta$ is observed to have 
a large enhancement in the presence of finite strangeness in the medium
due to additional contributions from the hyperons. The effects 
due to isospin asymmetry on the shear viscosity coefficient
is however observed to be marginal both in nuclear and hyperonic 
matter. 
The present study can be relevant for the experimental observables
of asymmetric relativistic heavy-ion collisions, e.g.,
in the compressed baryonic matter (CBM) experiment 
at the FAIR facility at GSI as well as in the future J-PARC-HI program.
\end{abstract}
\maketitle

\section{Introduction}
\label{intro}

The study of the thermodynamic and transport properties of 
QCD matter is an important area of research in contemporary 
strong interaction physics. The topic has relevance in heavy-ion collisions, as they can modify
the experimental observables, e.g., particle yields,
spectra, and collective flow and can give information
regarding the QCD phase transition. The transport properties have
been studied extensively in the literature in the confined 
\cite{Prakash_Phys_Rep_227_321_1993,Danielewicz_PLB_146_168_1984,Abu_Samreh_NPA552_1993_101,NPA573_554_1994_Monras_Trans_coeffs_Nucl_Neutron_matter,Hakim_Mornas_PRC47_2846_1993,A_S_Khvorostukhin_NPA915_198_2013,Itakura_PRD77_014014_2008,PRC86_024913_2012,PRC77_024911_2008_Gorenstein,
AD_HM_RM_PRD_106_2022_14013}
as well as in the quark-gluon plasma (QGP)  
\cite{Gavin_NPA435_1985_826,Hosoya_Kajantie_NPB250_666_1985,Phys_Rev_D31_53_1985_Danielewicz_Gyulassy,PRC84_035202_2011_visc_gluon_matter,A_S_Khvorostukhin_NPA845_106_2010} phases.
The hadronic transport approaches are used to describe
the dynamics of heavy ion collisions at low energies when the system
consists of hadrons.
However, for high-energy heavy-ion collisions, the hybrid 
transport approaches 
\cite{Hybrid_HIC,Hybrid_HIC_shear_1,Gotz_Hanah_PRC106_054904_2022,Gotz_Hanah_2503_10181}
are often used, which incorporate 
the relativistic viscous hydrodynamics for the QGP phase
along with the hadronic transport theory.
There are extensive studies in the 
literature of the transport coefficients within the hadron 
resonance gas model \cite{PRC86_024913_2012,PRC77_024911_2008_Gorenstein,
AD_HM_RM_PRD_106_2022_14013}, in which the hadrons are treated as
non-interacting particles. 
Within the Quantum Hadrodynamics (QHD) framework 
\cite{Recent_Prog_QHD_Serot_Walecka_IJMPE6_515_1996},
the interactions of the nucleons with the scalar and vector mesons
lead to medium modifications of the nucleons and their
effects on the transport coefficients are observed to be important.
%%%****************************************
The transport coefficients of hot nuclear matter have been studied
using the first-order Chapman--Enskog approximation with
the Boltzmann--Uehling--Uhlenbeck (BUU) collision term 
of the Boltzmann equation
\cite{NPA573_554_1994_Monras_Trans_coeffs_Nucl_Neutron_matter,BUU_5,BUU_6,BUU_7,BUU_8}
incorporating the Pauli blocking effects in Refs. \cite{Danielewicz_PLB_146_168_1984,Abu_Samreh_NPA552_1993_101}
as well as accounting for the  medium modifications of nucleons 
within the QHD framework. The results for the coefficients of viscosity
and thermal conductivity of hot nuclear matter in the Walecka model
obtained in a mathematically much simpler 
relaxation time approximation
\cite{Hakim_Mornas_PRC47_2846_1993} 
is observed to be very similar 
to the results obtained using the BUU collision term
\cite{NPA573_554_1994_Monras_Trans_coeffs_Nucl_Neutron_matter}.

The transport coefficients of viscosity and thermal conductivity
of hot nuclear matter have been derived 
incorporating the in-medium effects of the nucleons 
due to their interactions with the scalar and vector mesons within a quantum 
hadrodynamic framework in Ref.~\cite{Albright_Kapusta_PRC93_014903_2016}.
Recently, the transport coefficients of hot isospin asymmetric nuclear 
matter have been studied within a chiral SU(3) model, and 
the effects due to the medium modifications of the nucleons on 
the coefficients of shear viscosity and thermal conductivity are observed 
to be quite appreciable as compared to free nucleon gas
\cite{AM_JSB_shear_Kappa_HANM}.
The chiral SU(3) model
adopted in the present paper describes well the properties
of nuclear matter, finite nuclei, neutron stars
\cite{paper3,hartree,kristof1,Schramm_2013,Dex_2015,JSB_2016}
 and has also been 
used to study strange and heavy flavor mesons
\cite{AMSPM_EPJA_57_2021,AMSPMWG_2015,AMSPM_2017,AMSPM_DW_HQ_DS_PV_2023,AMAKSPM24}. 
In the chiral SU(3) model, the baryon masses are generated from spontaneous
chiral symmetry breaking, with the scalar mesons (proportional
to the light quark condensates) attaining nonzero
expectation values. In the hot  hadronic matter, the baryon masses
are obtained from the mean values of the scalar fields
($\sigma \sim (\bar u u + \bar d d)$, $\zeta \sim \bar s s$, 
$\delta \sim (\bar u u - \bar d d))$, which are obtained by 
solving their equations of motion. The negative mass shifts
of the baryons in the hadronic medium are thus associated with partial
restoration of the chiral symmetry in the chiral SU(3) model.
The chemical potentials of the baryons are modified
due to the interactions with the vector ($\omega$, $\rho$, and $\phi$) mesons
in the hot strange hadronic matter.
In the present work, using the chiral SU(3) model, 
we study the thermodynamic and transport properties of hadronic matter
comprising the nucleons and the hyperons.
The transport coefficients, namely, the diffusion coefficients
(corresponding to the multiple charges, namely the baryon number,
isospin, and the strangeness) as well as the
shear viscosity for strange hadronic matter are investigated
in the relaxation time approximation. 

The paper is organized as follows. In Sec.~\ref{sec:chiral-su3-model}, the chiral SU(3) model
used for the present study of the thermodynamic and transport properties
of strange hadronic matter is briefly described. Section \ref{sec:transport-coefficients} describes
the derivation of the diffusion matrix associated with multiple
charges (baryon number, isospin, and strangeness) as well as 
the coefficient of shear viscosity from the Boltzmann equation
using the first-order Chapman--Enskog expansion within the relaxation
time approximation. Section \ref{sec:results-discussion} describes the results obtained for the
thermodynamic and transport properties of strange hadronic matter, using the medium modifications of the baryon 
(antibaryon) within the chiral SU(3) model. Section \ref{sec:summary} summarizes the findings
of the present paper.

\section{Chiral SU(3) model}
\label{sec:chiral-su3-model}

In this section, we give a brief description of the chiral SU(3) model 
\cite{paper3,hartree,kristof1,AMAKSPM24} used in this work. Recently, the model has been used
to study the transport properties, e.g, the
shear viscosity and thermal conductivity
of hot nuclear matter \cite{AM_JSB_shear_Kappa_HANM}.
In the present work, we investigate the diffusion coefficients
arising due to multiple charges, e.g., baryon number, isospin
asymmetry, and strangeness charges, as well as 
shear viscosity of hot asymmetric strange hadronic matter.

The model \cite{paper3,hartree,kristof1}
is based on a nonlinear realization of chiral symmetry 
\cite{Weinberg,coleman, Bardeen} and 
incorporates the broken scale invariance
of QCD through a scalar dilaton field, $\chi$
\cite{sche1,heide1}. We use the mean field approximation,
in which the meson fields are treated as classical fields.
Also, for uniform and rotationally  invariant matter,
the scalar--isoscalar nonstrange $\sigma$, scalar--isoscalar
strange $\zeta$ and scalar--isovector $\delta$ fields are 
replaced by space-time independent mean values. For the vector fields, the spatial components
have zero expectation values, i.e, for isoscalar--vector fields,
$\omega^\mu \rightarrow \delta^{\mu 0}\omega^\mu
\equiv \delta^{\mu 0} \omega$, 
$\phi^\mu \rightarrow \delta^{\mu 0}\phi^\mu
\equiv \delta^{\mu 0} \phi$, and, for the isovector--vector field
$\rho^{\mu a} \rightarrow \delta^{\mu 0} \delta^{a 3} 
\equiv \delta^{\mu 0} \delta^{a 3} \rho$.
The meson fields which have non-zero contributions
to the baryon--meson interactions are the scalar and the vector mesons,
with the interaction Lagrangian given as
\begin{equation}
{\cal L}_{BS}+ {\cal L}_{BV}=\sum _a {\bar  \psi}^a
(g_{\sigma a} \sigma+g_{\zeta a} \zeta+g_{\delta a} \delta
-g_{\omega a}\gamma_\mu\omega^\mu
-g_{\rho a}\gamma_\mu {\vec \tau} \cdot {\vec {\rho^{\mu}}}
-g_{\phi a}\gamma_\mu\phi^\mu)\psi^a,
\label{BS_BV}
\end{equation}
where, $a$=$p$, $n$, $\Lambda$, $\Sigma^{\pm,0}$, and $\Xi^{-,0}$
refer to the nucleons and the hyperons.
The interaction of the baryons with the scalar and the vector
mesons leads to the effective mass and the effective chemical
potential of the $a$-th baryon to be given as
\begin{equation}
m_{a}^{*} = - g_{\sigma a}\sigma- g_{\zeta a}\zeta - g_{\delta a} \delta
\label{ameff}
\end{equation}
and 
\begin{equation}
\mu^*_a=\mu_a-g_{\omega a}\omega-g_{\rho a}\rho-g_{\phi a}\phi,
\label{amustr}
\end{equation}
respectively.

The Lagrangian density in the mean field approximation is
given as
\begin{equation}
{\cal L}={\cal L}_{BS}+ {\cal L}_{BV}+{\cal L}_{\mathrm{vec}}+{\cal L}_{0}
+{\cal L}_{\mathrm{scale}\mbox{-}\mathrm{break}}+{\cal L}_{\rm {SB}},
\label{L_MFT}
\end{equation}
with ${\cal L}_{BS}+{\cal L}_{BV}$ as given by Eq.~(\ref{BS_BV}), and
\begin{equation}
{\cal L}_{\mathrm{vec}} = \frac{1}{2} \frac{\chi^2}{\chi_0^2}\left(
m_{\omega}^{2} \omega^ 2+m_{\rho}^{2} \rho^ 2
+m_{\phi}^{2} \phi^ 2
\right) +g_4 \left(\omega^4 
+2 \phi^4
+6 \omega^2 \rho^2+\rho^4\right),
\label{L_vec}
\end{equation}
\begin{eqnarray}
{\cal L}_{0} & =& - \frac{ 1 }{ 2 } k_0 \chi^2
\left(\sigma^2+\zeta^2+\delta^2 \right) + k_1 \left(\sigma^2+\zeta^2+\delta^2 \right)^2
\nonumber \\
     &+& k_2 \left( \frac{ \sigma^4}{ 2 } + \frac{\delta^4}{2} + \zeta^4
 +3 \sigma^2 \delta^2 \right)
     + k_3 \chi \left(\sigma^2 - \delta^2 \right) \zeta - k_4 \chi^4, 
\label{L_0}
\end{eqnarray}
\begin{equation}
{\cal L}_{\mathrm{scale}\mbox{-}\mathrm{break}}= -\frac{1}{4} \chi^{4} 
{\rm ln} \frac{\chi^{4}}{\chi_{0}^{4}} + \frac{d}{3} \chi^{4} 
{\rm ln} \left[ \frac{\left( \sigma^{2} - \delta^{2}\right)\zeta }
{\sigma_{0}^{2} \zeta_{0}} \left( \frac{\chi}{\chi_{0}}\right) ^{3}\right],
\label{scalebreak}
\end{equation}
are the Lagrangian densities corresponding to
the vector mesons, scalar mesons, and the dilaton field,
with ${\cal L}_{\mathrm{scale}\mbox{-}\mathrm{break}}$ simulating the breaking of
scale invariance of QCD. ${\cal L}_{SB}$ is the explicit 
chiral symmetry breaking term given as
\begin{equation}
{\cal L} _{SB} =  - \left( \frac{\chi}{\chi_{0}}\right) ^{2} 
\left[ m_{\pi}^{2} 
f_{\pi} \sigma + \left( \sqrt{2} m_{K}^{2}f_{K} - \frac{1}{\sqrt{2}} 
m_{\pi}^{2} f_{\pi} \right) \zeta \right]. 
\label{l_ecsb}
\end{equation}
The pressure, $P$, which is the negative of the thermodynamic
potential per unit volume, is related to the thermodynamic quantities
such as temperature ($T$),
energy density ($\epsilon$), entropy density ($s$), chemical potential ($\mu_a$), and the number density ($\rho_{a}$) of the $a$-th baryon as
\begin{equation}
P=-\Omega/V=-\left(\epsilon-Ts-\sum_a \mu_a \rho_a \right),
\label{pressure}
\end{equation}
where,
\begin{eqnarray}
\epsilon 
&  = &  
\sum_a \gamma_a \int \frac{d^3 {\bf p}}{(2\pi)^3} 
E_a^*({\bf p})
\left[f_a^{\mathrm{\mathrm{eq}}}({\bf p})+{\bar f}_a^{\mathrm{\mathrm{eq}}}({\bf p})\right]
- {\cal L}_{\mathrm{vec}} - {\cal L}_0 
\nonumber \\
& - &  {\cal L}_{\mathrm{scale}\mbox{-}\mathrm{break}}
 - {\cal L}_{SB} -{\cal V}_{\mathrm{vac}} 
+\left(g_{\omega a}\omega+g_{\rho a}\rho \right) \rho_a,
\label{energy_density}
\end{eqnarray}
\begin{eqnarray}
s & = & 
-\sum_a \gamma_a \int \frac{d^3 {\bf p}}{(2\pi)^3} 
\left[f_a^{\mathrm{\mathrm{eq}}}({\bf p}) \ln f_a^{\mathrm{\mathrm{eq}}} ({\bf p})
+(1-f_a^{\mathrm{\mathrm{eq}}}({\bf p})) \ln (1- f_a^{\mathrm{\mathrm{eq}}} ({\bf p})) \right.\nonumber \\
&+& \left. {\bar f_a}^{\mathrm{\mathrm{eq}}}({\bf p}) \ln {\bar  f_a}^{\mathrm{\mathrm{eq}}} ({\bf p})
+(1-{\bar f_a}^{\mathrm{\mathrm{eq}}}({\bf p})) \ln (1- {\bar  f_a}^{\mathrm{\mathrm{eq}}} ({\bf p}))
\right],
\label{entr}
\end{eqnarray}
%and the number density of the $a$-th baryon, $\rho_a$ is 
%given by the expression
\begin{align}
\rho_{a}=  
\label{rho_a}
\gamma_a \int \frac{ d^3 {\bf p}}{(2\pi)^3} 
\left[f_a^{\mathrm{\mathrm{eq}}}({\bf p})-{\bar f}_a^{\mathrm{\mathrm{eq}}}({\bf p})
%\frac {1}{1+e^{E_{i}^*(k)-\mu^{*}_i)}/{T}}-
%\frac {1}{1+e^{(E_{i}^*(k)+\mu^{*}_i)}/{T}}
\right].
\end{align}
In the above equations,
$\gamma_a=2$ is the spin degeneracy factor and 
$E_a^*({\bf p}) =\left({{\bf p}^2+m_a^*}^2 \right)^{1/2}$
is the single-particle energy of the $a$-th baryon, 
with the effective mass and effective chemical
potential given by Eqs.~(\ref{ameff})
and (\ref{amustr}), respectively. 
The particle and antiparticle distribution functions 
in thermal equilibrium for the $a$-th baryon
%,$f_a^{\mathrm{\mathrm{eq}}} ({\bf p})$ and ${\bar f_a}^{\mathrm{\mathrm{eq}}} ({\bf p})$
are given as
\begin{equation}
f_a^{\mathrm{\mathrm{eq}}}({\bf p})= \frac {1}{e^{{\left(E_{a}^*({\bf p})-\mu^{*}_a \right)}/{T}} +1},\;\;
{\bar f_a}^{\mathrm{\mathrm{eq}}}({\bf p})
= \frac {1}{e^{{\left(E_{a}^*({\bf p})+\mu^{*}_a \right)}/{T}} +1}.
\label{equil_distr_fns}
\end{equation}
In Eq.~(\ref{energy_density}), the potential ${\cal V}_{\mathrm{vac}}(=-{\cal L}_{\mathrm{vac}})$ at $\rho_B=0$ and $T = 0$ 
has been subtracted to ensure vanishing vacuum energy.
The values of the scalar fields, $\sigma (\sim (\langle \bar u u \rangle +
\langle \bar d d \rangle))$ and $\zeta (\sim \langle \bar s s \rangle) $ in vacuum,
$\sigma_0$ and $\zeta_0$, are related to the 
pion and kaon decay constants, $f_\pi$ and $f_K$ as,
$\sigma_0=-f_\pi$
and $\zeta_0=-\frac{1}{\sqrt 2} (2 f_K -f_\pi)$, in accordance with
the partial conservation of axial current (PCAC) relations
for the pion and kaon \cite{paper3}.
The parameters $k_0$, $k_2$
and $k_4$ of the Lagrangian density for the scalar mesons, ${\cal L}_0$
are obtained by ensuring extrema in the vacuum for the
equations of motion of scalar fields, $\sigma$, $\zeta$ and the
dilaton field, $\chi$, obtained through minimization of the 
thermodynamic potential.
The parameters $k_1$ and $k_3$ are fitted to reproduce
the mass of $\sigma$ to be of the order of 500 MeV and the $\eta$ and $\eta'$ masses, respectively.
On the other hand, the value of
the $\chi$ in vacuum is fitted so that the pressure $P=0$ 
at the nuclear matter saturation density. The parameters used in 
the present study are the same as adopted in the recent study of the transport
properties of hot nuclear matter \cite{AM_JSB_shear_Kappa_HANM}.

For the hot strange hadronic matter, the input thermodynamic quantities are the temperature $T$, the baryon density $\rho_B = \sum_a \rho_a$, the isospin asymmetry parameter $t_A = \sum_a 2I_{3a}\rho_a / \rho_B$, and the strangeness fraction $f_s = \sum_a |s_a| \rho_a / \rho_B$. Here $I_{3a}$ and $s_a$ are the third isospin component and the number of strange quarks of the $a$-th baryon, respectively. Given these values, the mean-field solutions for the scalar fields ($\sigma$, $\zeta$, and $\delta$), the dilaton field $\chi$, and the vector fields ($\omega$, $\rho$, and $\phi$) are obtained by solving their coupled equations of motion, which follow from minimizing the thermodynamic potential.
The difference in the spatial density distributions
of the protons and neutrons of the colliding nuclei (due to 
$N\ne Z$) leads to isospin asymmetry in the hadronic matter,
and, in the present work, we choose the asymmetry parameter $t_A=0.6$
\cite{AM_JSB_shear_Kappa_HANM} as a typical value.

\section{Transport Coefficients}
\label{sec:transport-coefficients}

In the present paper, we investigate the diffusion and shear viscous coefficients of hot asymmetric 
strange hadronic matter using a chiral SU(3) model.
The components of the diffusion matrix corresponding to
multiple conserved charges: the baryon number, 
isospin, and strangeness are obtained  
from the Boltzmann equation, 
retaining the first-order deviations
of the baryon distribution functions 
from the local equilibrium distribution functions.
The coefficients of viscosity and thermal conductivity
of hot nuclear matter have been studied 
within a quasiparticle 
hadrodynamic framework in Ref.~\cite{Albright_Kapusta_PRC93_014903_2016}
and, recently, within the chiral SU(3) model
in Ref.~\cite{AM_JSB_shear_Kappa_HANM}.
In the present work, the diffusion matrix arising due to
the baryon, isospin, and strangeness charges,
as well as the coefficient of shear viscosity
of hot isospin asymmetric strange hadronic matter are 
studied within the relaxation time approximation. 
The baryons (nucleons and hyperons)
interact with the scalar mesons ($\sigma$, $\zeta$, and $\delta$) 
and the vector mesons ($\omega$, $\rho$, and $\phi$) described by
the Lagrangian density given by Eq.~(\ref{BS_BV}) 
within the chiral SU(3) model. 
For uniform hot hadronic matter in thermal and kinetic equilibrium, 
the scalar and vector fields have space-time independent 
mean values and the expectation values of the spatial components 
of the vector fields vanish due to rotational 
invariance in the rest frame of the hadronic medium
($u^\mu=(1,{\bf 0})$). However, for out-of-equilibrium nonuniform
matter with flow velocity $u^\mu=(1, {\bf v})$, the mean fields 
are space-time dependent, and the spatial
components of the vector fields no longer vanish.
The dispersion relation for the $a$-th baryon (antibaryon)
is then given as 
\begin{equation}
E_a ({\bf p}_a)=E_a^*({\bf p}^*_a)
\pm (g_{\omega a}{\bf \omega} +g_{\rho a}{\bf \rho}
+g_{\phi a}{\bf \phi}),
\label{energy_baryon_tot}
\end{equation}
where $E_a^* \left({\bf p}^*_a)=({{\bf p}^*_a}^2+{m_a^*}^2 \right)^{1/2}$
is the single particle energy given
in terms of the kinetic momentum,
\begin{equation}
{\bf p}^*_a={\bf p} \mp (g_{\omega a}
{ \mbox{\boldmath $\omega$}} 
+g_{\rho a} {\mbox{\boldmath $\rho$}}
+g_{\phi a} {\mbox{\boldmath $\phi$}}). 
\label{kin_mom}
\end{equation}
The energy-momentum tensor is given as
\begin{equation}
T^{\mu \nu}=-pg^{\mu \nu}+{\cal W} u^\mu u^\nu 
+\Delta T^{\mu \nu},
\label{energy-mom}
\end{equation}
where ${\cal W}=\epsilon +P $ is the enthalpy density and
\begin{equation}
\Delta T^{\mu \nu}=\eta \left(D^\mu u^\nu+D^\nu u^\mu
+\frac{2}{3}\Delta^{\mu \nu}\partial_\alpha u^\alpha\right)
-\xi \Delta^{\mu \nu}\partial_\alpha u^\alpha
\label{energy-mom_diss}
\end{equation}
is the dissipative part of the energy-momentum tensor,
$\eta$ and $\xi$ in $\Delta T^{\mu\nu}$ are the coefficients
of shear and bulk viscosity.

The currents corresponding to conserved charge, $q_i$
for $i=1$, $2$, and $3$, associated with the baryon number, isospin, 
and strangeness, 
are given by
\begin{equation}
J_{q_i}^{\mu}=\rho_{q_i} u^\mu +\Delta J_{q_i}^\mu,
\label{q_i_current}
\end{equation}
with the dissipative part 
\begin{equation}
\Delta J_{q_i}^{\mu}=\kappa_{ij}
D^\mu\left(\frac{\mu_{q_j}}{T}\right),
\label{q_i_current_diss}
\end{equation}
given in terms of the diffusion coefficients, 
$\kappa_{ij} (\equiv \kappa_{q_i q_j})$,
$ij$-elements of the diffusion matrix $\kappa$.
In the above equations, 
$D^\mu=\partial ^\mu -u^\mu D$, 
$D=u^\alpha \partial _\alpha$,
$\Delta ^{\mu \nu}=u^\mu u^\nu -g^{\mu \nu}$.
The coefficients of viscosity and diffusion 
are computed from the dissipative parts of the energy-momentum tensor 
and the current, $\Delta T^{\mu \nu}$ and
$\Delta J_{q_i}^\mu$ respectively. 

The Boltzmann equation for the distribution functions 
of baryon (antibaryon) of species $a$=($p$, $n$, $\Lambda$, $\Sigma^{\pm,0}$,
and $\Xi^{-,0})$ for the strange hadronic matter as considered 
in the present paper is given as
\begin{equation}
\frac{ d {F_a({\bf x}, {\bf p}^*_a, t)}}{dt}=C_a,
\label{Boltzmann_eqn}
\end{equation}
where $F_a=f_a$ $(\bar f_a)$ is the distribution function
of the particle (antiparticle).
Considering a first-order deviation from local equilibrium
distribution function, $F_a$ is written in terms of 
function $\phi_a (x,p)$ as 
\begin{equation}
{F_a}=F_a ^{\mathrm{\mathrm{eq}}} (1+\phi_a (x,p)),
\end{equation}
where $F_a^{\mathrm{\mathrm{eq}}}$=${f_a} ^{\mathrm{\mathrm{eq}}}$ (${\bar {f_a}} ^{\mathrm{\mathrm{eq}}}$) 
is the distribution function of the $a$-th baryon (antibaryon)
in local equilibrium. 
The function $\phi_a(x,p)$ is expressed in the same 
tensorial form as the dissipative parts of the energy-momentum tensor
and the current corresponding to charge $q_i$ (given by
Eqs.~(\ref{energy-mom_diss}) and 
(\ref{q_i_current_diss}) respectively) as
\begin{equation}
\phi_a(x,p)={\tilde A}_a \partial_\alpha u^\alpha
-\sum_{q_i} {\tilde B}_a^{q_i} p^\mu_{a} D_\mu
\left(\frac{\mu_{q_i}}{T}\right)
+{\tilde C}_a p^\mu_a p^\nu_a \left(D_\mu u_\nu+D_\nu u_\mu
+\frac{2}{3}\Delta_{\mu \nu}\partial_\alpha u^\alpha\right),
\label{phi_a}
\end{equation}
where ${\tilde A}_a$, ${\tilde B}_a^{q_i}$, and ${\tilde C}_a$
are functions of $p$.
In the relaxation time approximation,
the collision integral on the R.H.S. of the Boltzmann equation 
assumes the form
\begin{equation} 
C_a=-\frac{F_a^{\mathrm{eq}}\phi_a}{\tau_a(E_a^*)},
\label{Collision_integral}
\end{equation}
where $\tau_a (E_a^*)$ is the relaxation time.
This leads to the form of the Boltzmann equation
(\ref{Boltzmann_eqn}) to be given as
\begin{equation}
\frac{d{F_a({\bf x}, {\bf p}^*_a, t)}}{dt}
=-\frac{F_a^{\mathrm{eq}}\phi_a}{\tau_a(E_a^*)},
\label{Boltzmann_eqn_relaxation_time}
\end{equation}
in the relaxation time approximation.
The left-hand side of the Boltzmann equation 
$\frac{d{F_i({\bf x}, {\bf p}^*_i, t)}}{dt}
\left(=\frac{\partial F_i}{\partial t}
+\frac{d {x^k}}{dt} \frac{\partial F_i}{\partial {x^k}} 
+\frac{d {{p_i^*}^k}}{dt} \frac{\partial F_i}{\partial {{p_i^*}^k}} \right)$,
is evaluated by using the local equilibrium form 
of the distribution function, $F_i^{\mathrm{eq}}({\bf x}, {\bf p}^*_i,t)$
\cite{Albright_Kapusta_PRC93_014903_2016,AM_JSB_shear_Kappa_HANM},
which acts as a source term 
for the collision term on the R.H.S. of the Boltzmann equation.
Equating the tensor structures on both sides
of the above equation, the particular solutions for the 
functions ${\tilde A}_a$,  ${\tilde B}^{q_i}_a$, and ${\tilde C}_a$ 
of the function $\phi_a(x,p)$ given by Eq.~(\ref{phi_a})
are obtained.
The left-hand side of
Eq.~(\ref{Boltzmann_eqn_relaxation_time}),
retaining only the part corresponding to the diffusion coefficients (generalized to multiple charges)
is given by  
\begin{equation}
\frac{d F_a^{\mathrm{eq}}}{dt}=\sum_{q_i}
\left(q_{ia} -\frac{\rho_{q_i} E_a}{{\cal W}}\right)
\frac{p^\mu_a}{E^*_a}D_\mu \left(\frac{\mu_{q_i}}{T}\right)  
F_a^{\mathrm{eq}},
\label{Boltzmann_eqn_LHS}
\end{equation}
where for $i=1$, $2$, and $3$, $q_{ia} \equiv b_a$ (baryon number), 
$i_a$ ($=2I_{3a}$), and $s_a$ (number of strange quarks)
are the baryon (antibaryon) charge of species $a$, 
associated with the baryon number, 
isospin, and strangeness, respectively, and, $\mu_{q_i}\equiv \mu_B$, $\mu_I
$, and $\mu_S$ is the corresponding chemical potential. 

Equating the L.H.S. of the Boltzmann equation (\ref{Boltzmann_eqn_LHS}) to the diffusion term
on the right-hand side
of Eq.~(\ref{Boltzmann_eqn_relaxation_time})
gives the particular solutions for the functions
${\tilde B}^{q_i}_a$ as
\begin{equation}
{{\tilde B}_a^{q_i\,(\mathrm{par})}}=\frac{\tau_a}{E^*_a}
\left(q_{ia}- \frac{\rho_{q_i} E_a}{{\cal W}}\right).
\label{B_a_qi_par}
\end{equation}
However, there is an arbitrariness in the solution of
${\tilde B}^{q_i}_a$, i.e., for a particular 
solution ${{\tilde B}_a^{q_i\,(\mathrm{par})}}$, 
one can generate another solution 
${\tilde B}^{q_i\,(\mathrm{par})}_a-b^{q_i}$, where $b^{q_i}$ is a constant
independent of the particle species.
This arbitrariness is resolved by imposing the Landau--Lifshitz 
condition, which requires $\delta T^{0k}$=0
in the local rest frame. This condition implies that
\begin{equation}
\delta T^{0k}=\sum_a \int d {\Gamma}_a \frac{{p^*_a}^k}{E^*_a}
E_a  \delta \tilde{F_a}=0,
\end{equation}
where, $d{\Gamma}_a=\gamma_a \frac{d^3 {{\bf p}^*_a}}{(2\pi)^3}$ and
$\delta \tilde{F}_a=F_a-F_a^{\mathrm{eq}}=F_a^{\mathrm{eq}}\phi_a$.
For the diffusion terms, the above condition yields
\begin{equation}
\delta T^{0k}=\sum_a  \int d {\Gamma}_a \frac{{p^*_a}^k}{E^*_a}
E_a  \left[-\sum_{q^i}({\tilde B}^{q_i}_a-b^{q_i}) {p_a^*}^\alpha 
D_\alpha \left(\frac{\mu_{q_i}}{T}\right)\right] F^{\mathrm{eq}}_a
=0.
\end{equation}
In the above integral, the temporal component of ${p_a^*}^\alpha$ 
yields zero as the integrand becomes an odd function of the
spatial component, ${p_a^*}^k$, and, using 
the spatial momentum isotropy ${p_a^*}^k {p_a^*}^l
\rightarrow \frac{1}{3}\delta^{kl}|{\bf p}_a^*|^2$, we obtain,
\begin{equation}
b^{q_i}\sum_a  \int d {\Gamma}_a \frac{|{{\bf p}^*_a}|^2}{E^*_a}
E_a  F^{\mathrm{eq}}_a
= \sum_a  \int d {\Gamma}_a \frac{|{{\bf p}^*_a}|^2}{E^*_a}
E_a {\tilde B}^{q_i}_a F^{\mathrm{eq}}_a,
\end{equation}
for each of the conserved charges,
$q_{ia}\equiv b_a$, $i_a$, and $s_a$ for baryon (antibaryon)
of species $a$. 
Using
\begin{equation}
\sum_a  \int d {\Gamma^*}_a \frac{|{{\bf p}^*_a}|^2}{E^*_a}
E_a  F^{\mathrm{eq}}_a=3 T{\cal W},
\end{equation}
we obtain the expression for $b^{q_i}$ as
\begin{equation}
b^{q_i}
=\frac{1}{3T{\cal W}} \sum_a  \int d \Gamma_a \frac{|{{\bf p}^*_a}|^2}{E^*_a}
E_a {\tilde B}^{q_i}_a F^{\mathrm{eq}}_a.
\label{b_q_i}
\end{equation}
Equating the spatial component of the dissipative part 
of the current associated with charge $q_i$, 
$\Delta J_{q_i}^{k}$ given by Eq.~(\ref{q_i_current_diss})
with the diffusion part of $\delta J_{q_i}^k$ yields
\begin{eqnarray}
\Delta J_{q_i}^k &=&
\sum_{q_j} \kappa_{q_i q_j} D^k\left(\frac{\mu_{q_j}}{T}\right)
= \left({\delta {J_{q_i}^k}}\right)_{\rm diffusion}\nonumber \\
&=&\sum_a q_{ia} \int d \Gamma_a 
\frac{{p_a^*}^k}{E_a^*}
\left( {\delta \tilde{F}_a}\right)_{\rm diffusion}
\nonumber \\
&=&\sum_a q_{ia} \int d \Gamma_a 
\frac{{p_a^*}^k}{E_a^*}
\left[-\sum_{q_j} {\tilde B}_a^{q_j} p_{ak'} D^{k'}
\left(\frac{\mu_{q_j}}{T}\right)\right] F_a^{\mathrm{eq}}.
\end{eqnarray}
Using $-p_{ak'}=p_a^{k'}={p_a^*}^{k'}$ in the local rest frame
and ${p_a^*}^k {p_a^*}^{k'}
\rightarrow \frac{1}{3}\delta^{kk'}|{\bf p}_a^*|^2$, 
we obtain the diffusion coefficients to be given as
\begin{eqnarray}
\kappa_{ij} \equiv \kappa_{q_i q_j} =\sum_a q_{ia} \int d \Gamma_a 
\frac{|{{\bf p}_a^*}|^2}{3E_a^*}
{\tilde B}_a^{q_j}
F_a^{\mathrm{eq}}.
\end{eqnarray}
Using the solution ${\tilde B}_a^{q_j}={\tilde B}_a^{q_j\,(\mathrm{par})}-b^{q_j}$
in the above equation, we obtain
\begin{eqnarray}
\kappa_{ij}& =& \sum_a q_{ia} \int d \Gamma_a 
\frac{|{p_a^*}|^2}{3E_a^*}
{\tilde B}_a^{q_j\,(\mathrm{par})} F_a^{\mathrm{eq}}
- b^{q_j} \sum_a q_{ia} \int d \Gamma_a 
\frac{|{p_a^*}|^2}{3E_a^*} F_a^{\mathrm{eq}}.
\end{eqnarray}
In the above equation, 
using the expression of $b^{q_j}$ given by Eq.~(\ref{b_q_i}),
and the factor multiplying $b^{q_j}$
to be given by
\cite{AD_HM_RM_PRD_106_2022_14013} 
\begin{eqnarray}
\sum_a q_{ia} \int d \Gamma_a 
\frac{|{{\rm p}_a^*}|^2}{3E_a^*} F_a^{\mathrm{eq}}=\rho_{q_i}T,
\end{eqnarray}
we obtain
\begin{eqnarray}
\kappa_{ij}& =& \sum_a 
 \int d \Gamma_a 
\frac{|{{\bf p}_a^*}|^2}{3E_a^*}
\left(q_{ia} -\frac{\rho_{q_i}}{{\cal W}}E_a\right)
{\tilde B}_a^{q_j\,(\mathrm{par})} F_a^{\mathrm{eq}}.
\end{eqnarray}
Using the particular solutions given by Eq.~(\ref{B_a_qi_par}),
the expressions for the diffusion coefficients, 
$\kappa_{ij}$, are obtained as
\begin{eqnarray}
\kappa_{ij} & = &
\sum_{a} \int d\Gamma_a 
\frac{|{\bf p}^*_a|^2}{3{E_a^*}^2}\tau_a (E_a^*) 
\left(q_{ia} -\frac{\rho_{q_i}}{{\cal W}} 
E_a \right) 
\left(q_{ja} -\frac{\rho_{q_j}}{{\cal W}} 
E_a \right) {F_a}^{\mathrm{eq}},
\label{kappa_ij}
\end{eqnarray}
where $E_a=E_a^*\pm 
(g_{\omega a}\omega +g_{\rho a}\rho +g_{\phi a}\phi)$
for the baryon (antibaryon). For $i=1$, $2$, and $3$, 
$\rho_{q_i}$ are the charge densities associated with the baryon number, 
isospin, and strangeness, respectively. These are given as 
$\rho_B \left(=\sum_a b_a \rho_a \right)$, $\rho_I \left(=\sum_a i_a \rho_a \right)$, 
and $\rho_S \left(=\sum_a s_a \rho_a\right)$, where $q_{ia} \equiv b_a$, $i_a$, and $s_a$ 
are the corresponding conserved charges.

The coefficient of shear viscosity, $\eta$, was studied
in hot nuclear matter in Ref.
\cite{AM_JSB_shear_Kappa_HANM}
within the chiral SU(3)
model. In the present work, the effect on $\eta$ due to 
the inclusion of hyperons is studied. 
The expression for the coefficient of shear viscosity 
is given by \cite{Albright_Kapusta_PRC93_014903_2016}
\begin{equation}
\eta=\frac{1}{15T}\sum_{a} 
\int  d\Gamma_a
\frac{{|{\bf p}^*_a|}^4}{{E_a^*}^2}
\tau_a \left(E_a^* \right) F_a^{\mathrm{eq}}
\label{eta}
\end{equation}
As was done in the previous study
\cite{AM_JSB_shear_Kappa_HANM},
the energy-dependent relaxation time for
the $a$-th baryon in the integrand of the expression
of the coefficient of shear viscosity is replaced 
by a medium-dependent mean value calculated 
from its average velocity, $\langle v_a \rangle$, and 
the mean free path, $\lambda_a$, using the formula
$\tau_a=\lambda_a/{\langle v_a \rangle}$.
For the diffusion and the shear viscosity coefficients, 
which correspond to the net current flow 
(baryon number, isospin, or strangeness)
and the energy-momentum flow, respectively,
the relaxation times are obtained
using the following expressions. 
For the diffusion coefficients,
\begin{equation}
\tau_a^d=\lambda_a^d/{\langle v_a^d \rangle},\,\,
\langle v_a^d\rangle
={\frac{1}{\rho_a}} 
\int d \Gamma_a \frac{|{\bf p}^*_a|}{E_a^*({\bf p}^*_a)} 
\left[f_i^{\mathrm{eq}}({\bf p}^*_a)-{\bar f}^{\mathrm{eq}}({\bf p}^*_a)
\right],\;\;\;
\lambda_a^{d} =1/\left(\rho_a \sigma_{BB}\right),
\label {v_almb_a_d}
\end{equation}
with $\rho_a= \int d \Gamma_a 
\left(f_a^{\mathrm{eq}}({\bf p}^*_a)-{\bar f}^{\mathrm{eq}}({\bf p}^*_a)\right)$
as the net baryon density of the $a$-th baryon,
whereas for the shear viscosity we take
\begin{equation}
\tau_a^s=\lambda_a^s/{\langle v_a^s \rangle},\,\,
\langle v_a^{s}\rangle
={\frac{1}{n_a^{tot}}} 
\int d \Gamma_a \frac{|{\bf p}^*_a|}{E_a^*({\bf p}^*_a)} 
\left[f_a^{\mathrm{eq}}({\bf p}^*_a)+{\bar f}^{\mathrm{eq}}({\bf p}^*_a)
\right],\;\;\;
\lambda_a^{s} =1/\left(n_a^{\mathrm{tot}} \sigma_{BB}\right),
\label {v_almb_a_shear}
\end{equation}
with $n_a^{\mathrm{tot}}=\int d \Gamma_a
\left(f_a^{\mathrm{eq}}({\bf p}^*_a)+{\bar f}^{\mathrm{eq}}({\bf p}^*_a)\right)$,
which is the sum of the number densities
of the $a$-th baryon and antibaryon.
In the above, $\sigma_{BB}$ is the total baryon--baryon
cross-section in vacuum, which, in the present work, is
taken to be the same as the nucleon--nucleon
cross-section in vacuum $\sigma_{NN} \sim 
\pi r_0^2$ (with
$r_0\sim 1.12$ fm as the radius of the nucleon), yielding
a value of $40$ mb 
\cite{Danielewicz_PLB_146_168_1984,Itakura_PRD77_014014_2008}.
This follows from the assumption
that the radius of the hyperon is similar to the radius
of the nucleon.

Using the expressions for the medium-dependent
ralaxation times given by Eqs.~(\ref{v_almb_a_d})
and  (\ref{v_almb_a_shear}),
the coefficients of diffusion and shear viscosity 
given by Eqs.~(\ref{kappa_ij}) and (\ref{eta}) can be re-written as
\begin{eqnarray}
\kappa_{ij} & = &
\sum_{a} \tau_a^d
\int d\Gamma_a 
\frac{|{\bf p}^*_a|^2}{3{E_a^*}^2}
\left(q_{ia} -\frac{\rho_{q_i}}{{\cal W}} 
E_a \right) 
\left(q_{ja} -\frac{\rho_{q_j}}{{\cal W}} 
E_a \right) {F_a}^{\mathrm{eq}},
\label{kappa_ij_vd}
\end{eqnarray}
where $E_a=E_a^*\pm 
(g_{\omega a}\omega +g_{\rho a}\rho +g_{\phi a}\phi)$
for the $a$-th baryon (antibaryon),
and,
\begin{equation}
\eta=\frac{1}{15T}\sum_{a} \tau_a^s 
\int  d\Gamma_a
\frac{{|{\bf p}^*_a|}^4}{{E_a^*}^2}
F_a^{\mathrm{eq}}.
\label{eta_vs}
\end{equation}
The coefficient of shear viscosity was already studied
for asymmetric nuclear matter in Ref.~\cite{AM_JSB_shear_Kappa_HANM}
within the chiral SU(3) model. Hence the effect of the strangeness 
is studied in the present paper.

\begin{figure}
\vskip -1.5in
\includegraphics[width=17cm,height=18cm]{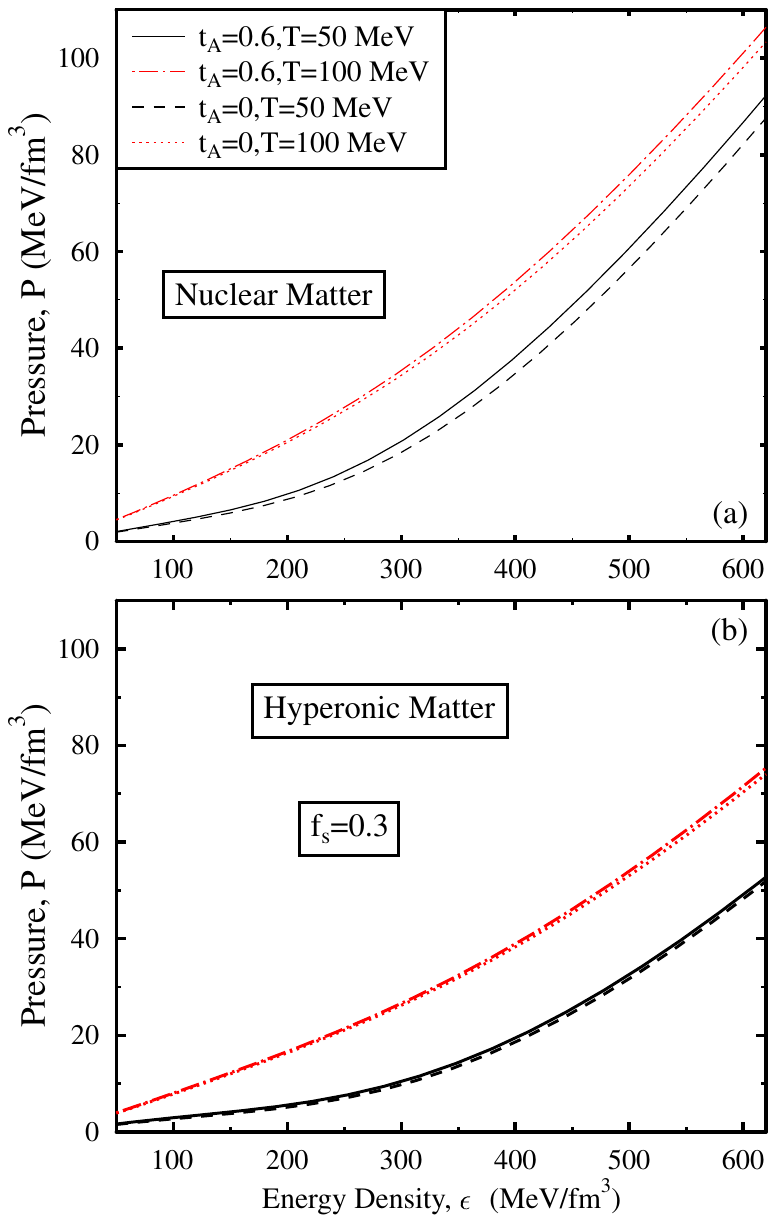} 
\vskip -0.1in
\caption{Pressure, $P$ (in MeV/fm$^3$),
is plotted as a function of energy density, $\epsilon$ (in MeV/fm$^3$),
for different values of temperature for isospin symmetric 
($t_A = 0$) and asymmetric matter (with asymmetry parameter, 
$t_A = 0.6$) in nuclear matter and hyperonic matter (with $f_s = 0.3$).}
\label{Dens_EOS_SHM}
\end{figure}

\begin{figure}
\vskip -1.5in
\includegraphics[width=17cm,height=18cm]{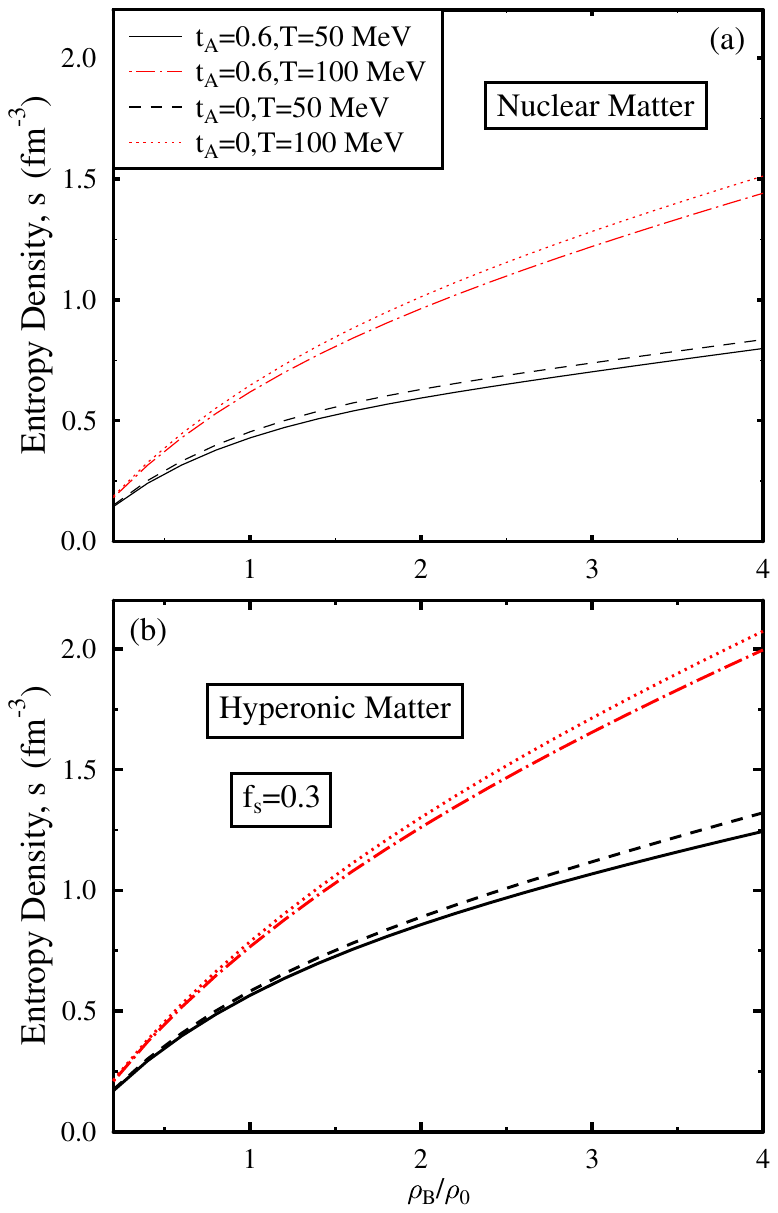} 
\vskip -0.1in
\caption{Entropy density, $s$ (in fm$^{-3}$),
is plotted as a function of the baryon density in units 
of nuclear matter saturation density, $\rho_B/\rho_0$, 
for different values of temperature for isospin symmetric 
($t_A = 0$) and asymmetric matter (with asymmetry parameter, 
$t_A = 0.6$) in nuclear matter and hyperonic matter (with $f_s = 0.3$).}
\label{Dens_Entr_SHM}
\end{figure}

\begin{figure}
\vskip -1.5in
\includegraphics[width=18cm,height=19cm]{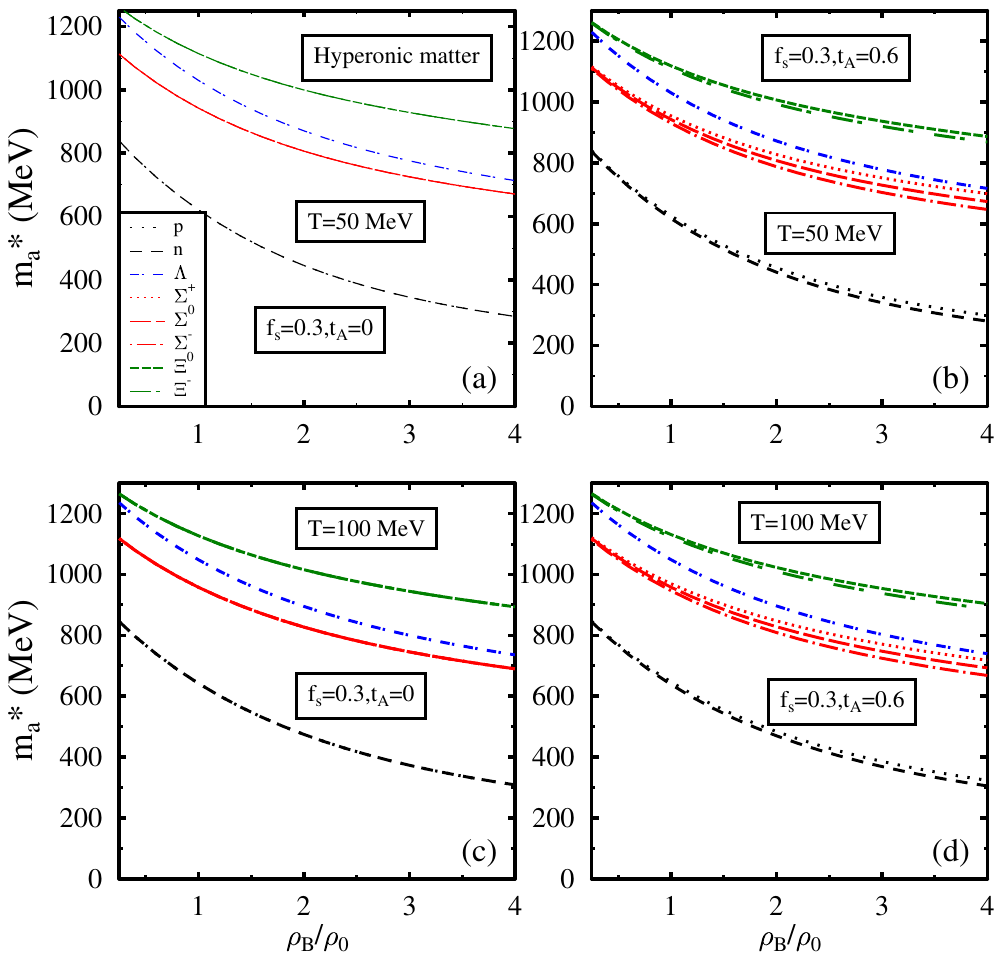} 
\vskip -0.3in
\caption{Effective masses, $m_a^*$ (in MeV), 
for $i=p$, $n$, $\Lambda$, $\Sigma^\pm$, $\Sigma^0$, and $\Xi^{0,-}$ are
plotted as functions of  the baryon density in units 
of nuclear matter saturation density, $\rho_B/\rho_0$,
for different values of temperature for isospin symmetric 
($t_A = 0$) and asymmetric (with asymmetry parameter, 
$t_A = 0.6$) hyperonic matter (with $f_s = 0.3$).}
\label{Dens_meffa}
\end{figure}

\begin{figure}
\vskip -1.5in
\includegraphics[width=18cm,height=19cm]{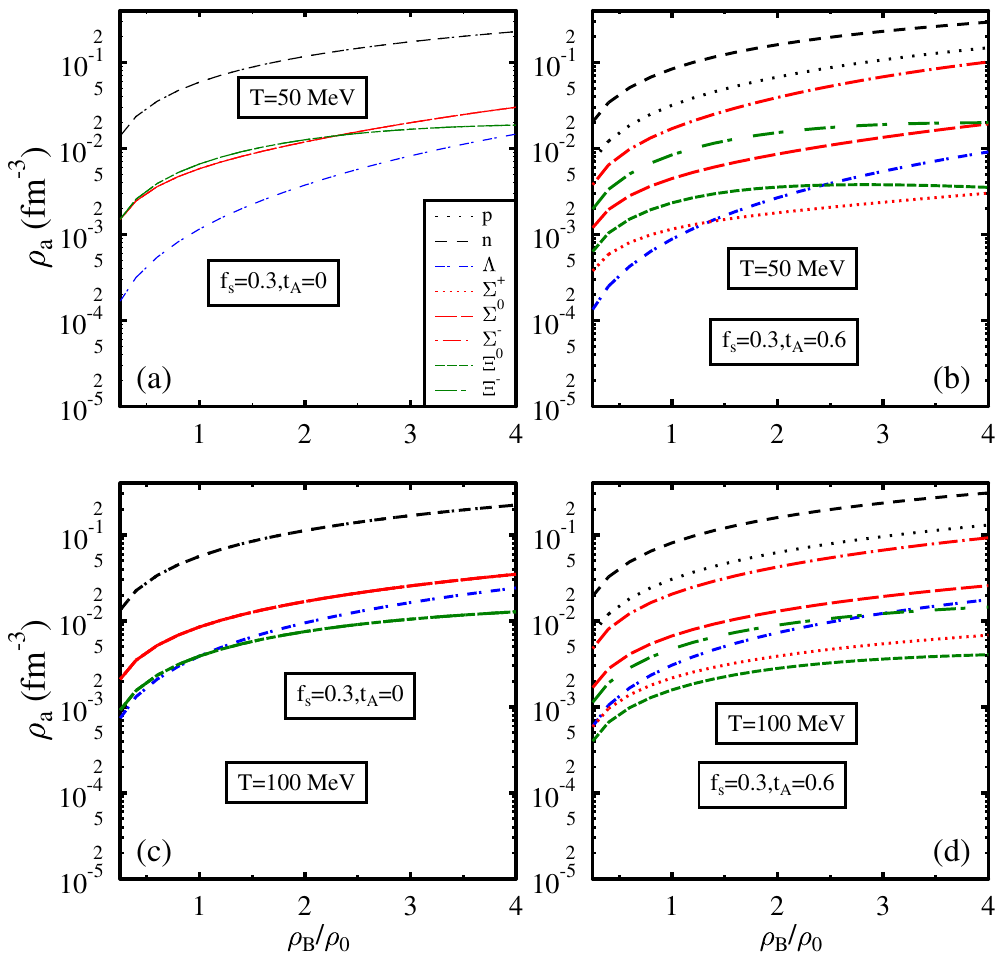} 
\vskip -0.3in
\caption{Number densities, $\rho_a$ (in fm$^{-3}$),
with $a=p$, $n$, $\Lambda$, $\Sigma^\pm$, $\Sigma^0$, and $\Xi^{0,-}$ are
plotted as functions of the baryon density in units 
of nuclear matter saturation density, $\rho_B/\rho_0$,
for different values of temperature for isospin symmetric 
($t_A = 0$) and asymmetric (with asymmetry parameter, 
$t_A = 0.6$) hyperonic matter (with $f_s = 0.3$).}
\label{Dens_rhoa}
\end{figure}

\begin{figure}
\vskip -1.5in
\includegraphics[width=18cm,height=19cm]{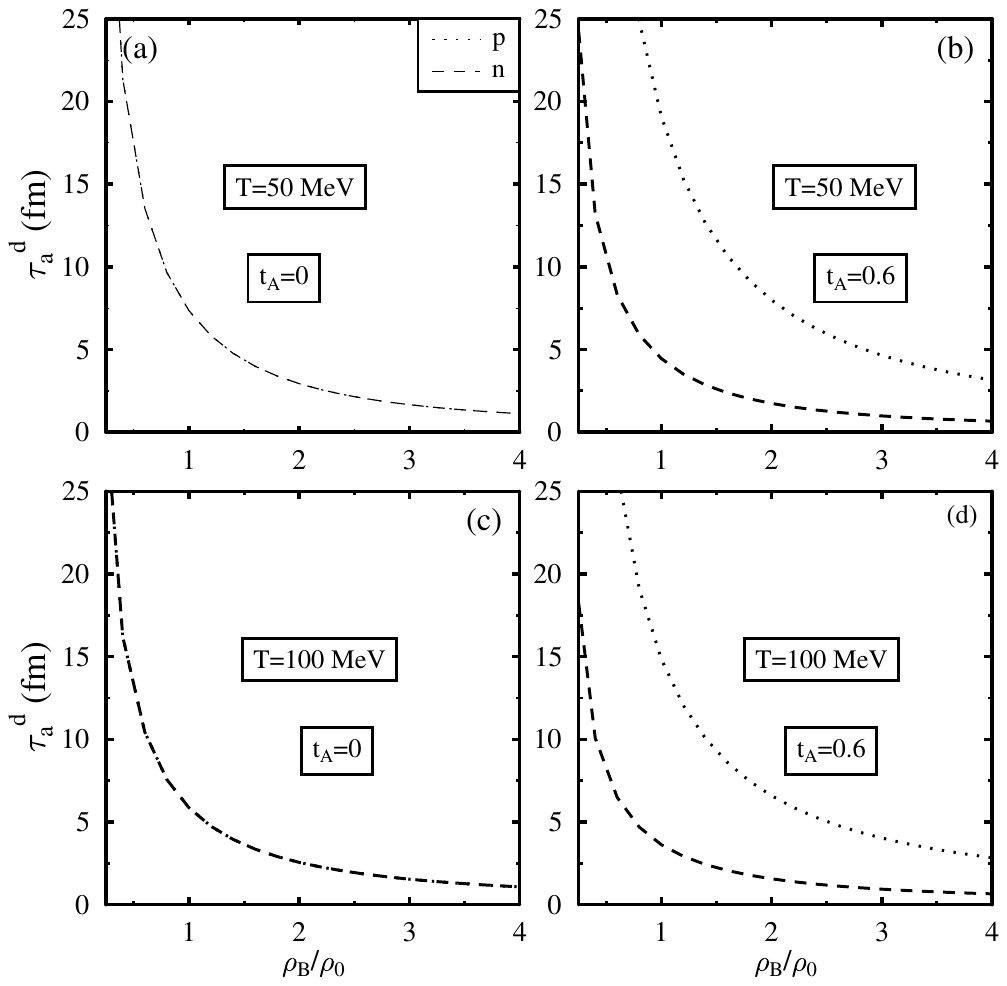} 
\vskip -0.7in
\caption{Relaxation times of the $a$-th nucleon, $\tau_a$ (in fm),
($a\equiv p$ or $n$) 
are plotted as functions of the baryon density 
in units of nuclear matter saturation density, $\rho_B/\rho_0$,
for different values of temperature for isospin symmetric 
($t_A = 0$) and asymmetric matter (with asymmetry parameter, 
$t_A = 0.6$) in nuclear matter.}
\label{Dens_tau_NM}
\end{figure}

\begin{figure}
\vskip -1.5in
\includegraphics[width=18cm,height=19cm]{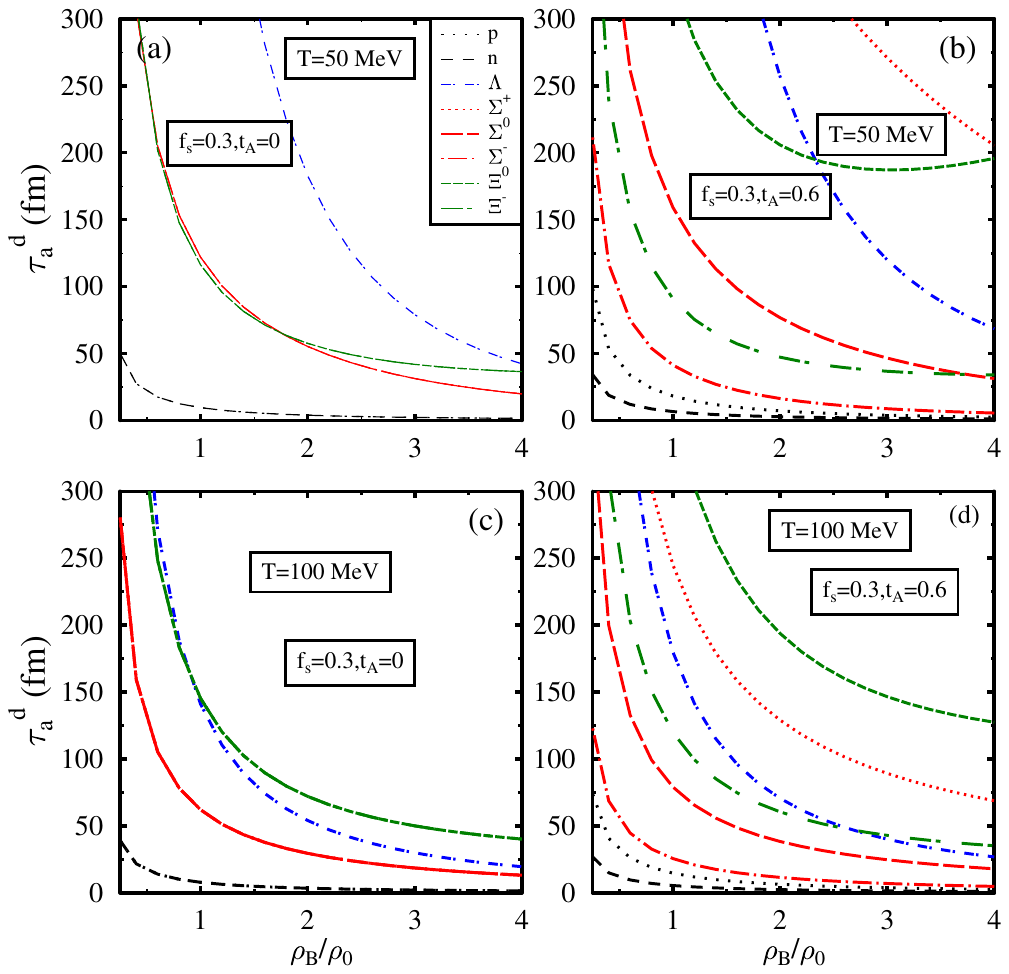} 
\vskip -0.3in
\caption{Relaxation times, $\tau_a^d$ (in fm), 
for $a=p$, $n$, $\Lambda$, $\Sigma^\pm$, $\Sigma^0$, and $\Xi^{0,-}$ are
plotted as functions of the baryon density in units 
of nuclear matter saturation density, $\rho_B/\rho_0$,
for different values of temperature for isospin symmetric 
($t_A = 0$) and asymmetric (with asymmetry parameter, 
$t_A = 0.6$) hyperonic matter (with $f_s = 0.3$).}
\label{Dens_tau_Nuc_Hyp}
\end{figure}

\begin{figure}
\vskip -1.5in
\includegraphics[width=18cm,height=19cm]{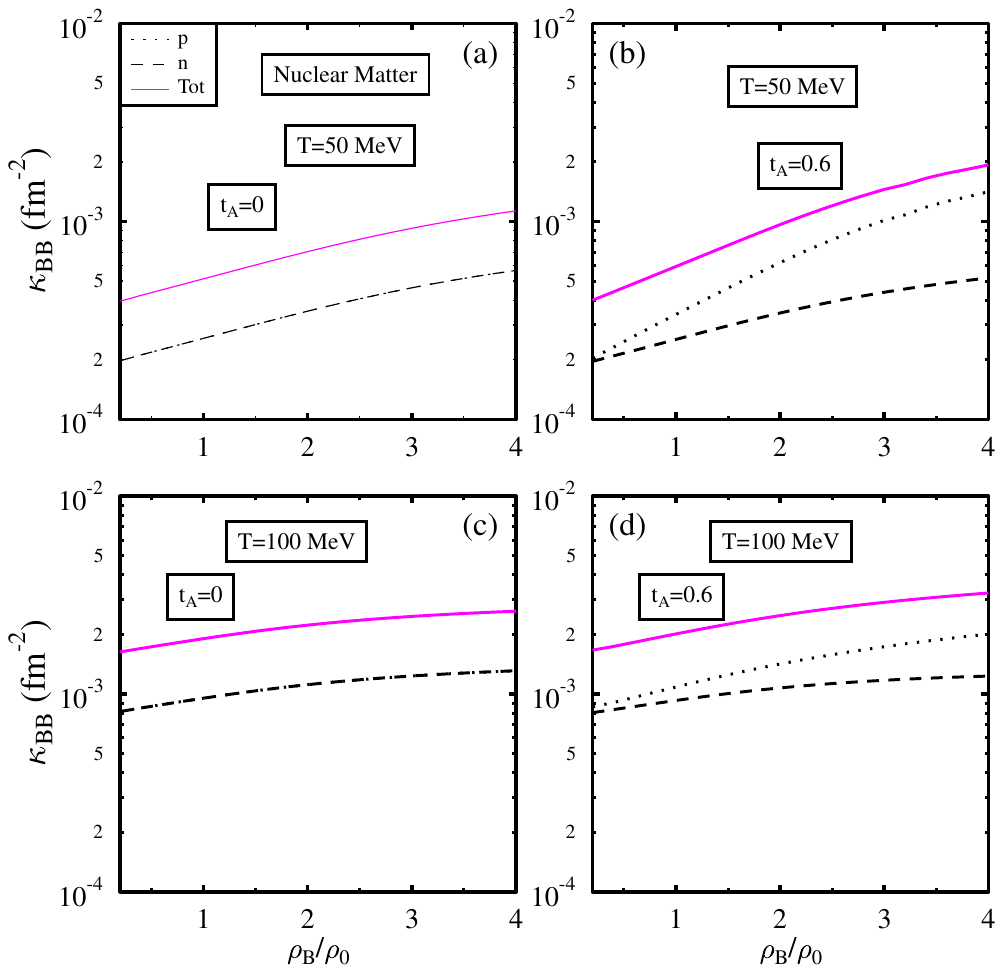} 
\vskip -0.5in
\caption{Diffusion coefficient, $\kappa_{BB}$ (in fm$^{-2}$),
is plotted as a function of the baryon density in units 
of nuclear matter saturation density, $\rho_B/\rho_0$, 
for different values of temperature for isospin symmetric 
($t_A = 0$) and asymmetric (with asymmetry parameter, 
$t_A = 0.6$) in nuclear matter.}
\label{Dens_kbb_ind_NM}
\end{figure}

\begin{figure}
\vskip -1.5in
\includegraphics[width=18cm,height=19cm]{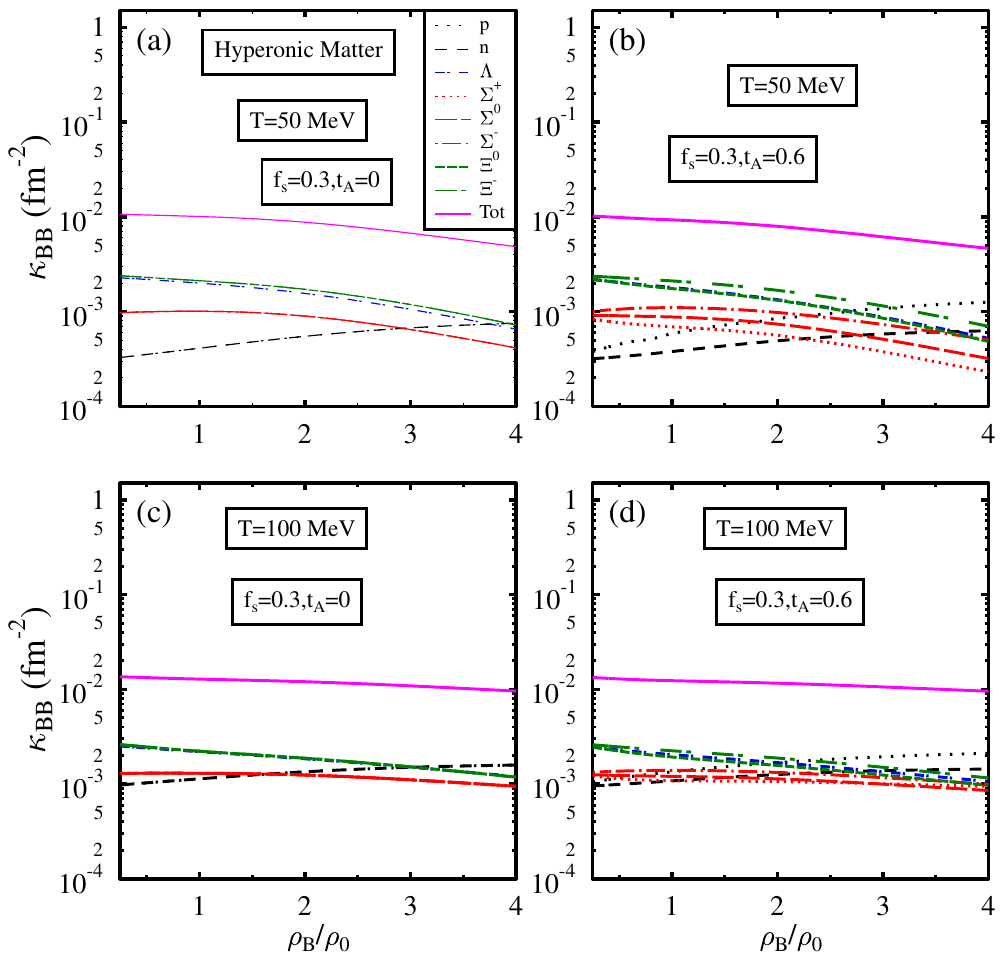} 
\vskip -0.5in
\caption{Diffusion coefficient, $\kappa_{BB}$ (in fm$^{-2}$),
is plotted as a function of the baryon density in units 
of nuclear matter saturation density,  $\rho_B/\rho_0$,
for different values of temperature for isospin symmetric 
($t_A = 0$) and asymmetric (with asymmetry parameter, 
$t_A = 0.6$) hyperonic matter (with $f_s = 0.3$).}
\label{Dens_kbb_ind_SHM}
\end{figure}

\begin{figure}
\vskip -1.5in
\includegraphics[width=18cm,height=19cm]{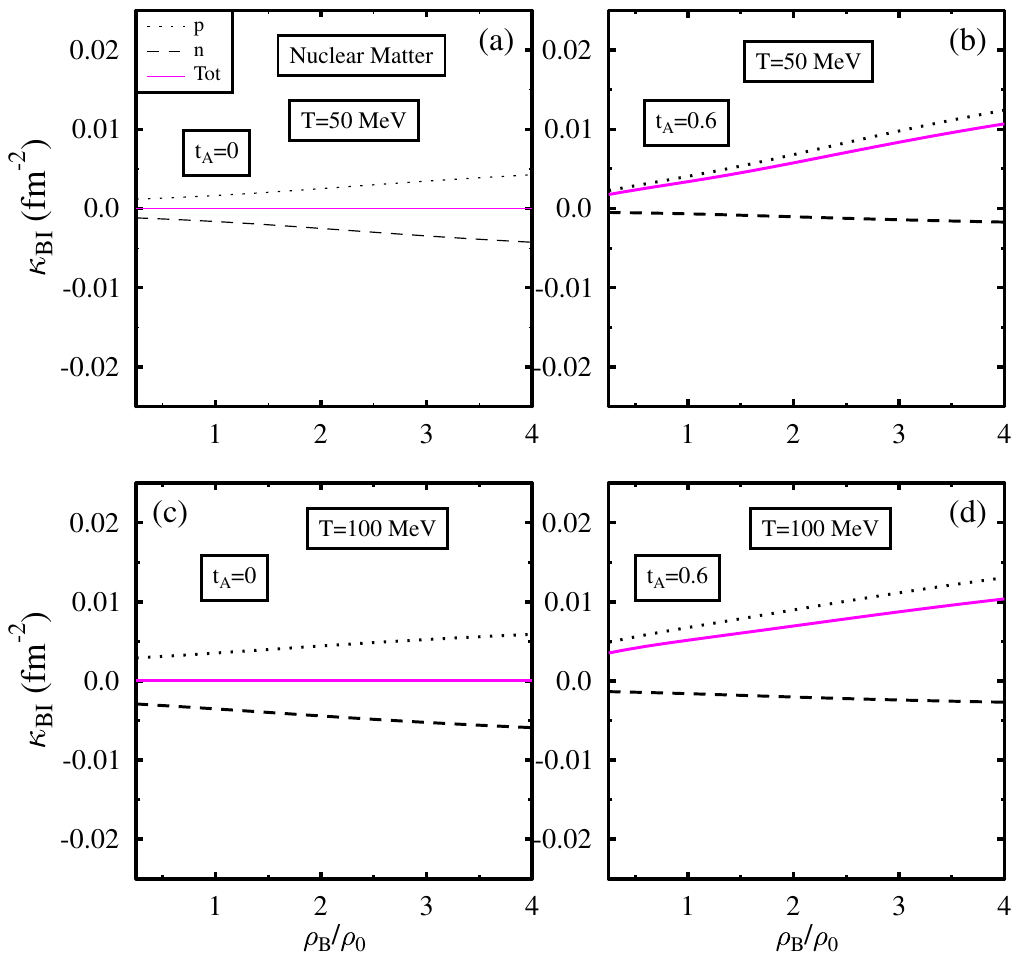} 
\vskip -0.5in
\caption{Diffusion coefficient, $\kappa_{BI}$ (in fm$^{-2}$),
is plotted as a function of the baryon density in units 
of nuclear matter saturation density,  $\rho_B/\rho_0$,
for different values of temperature for isospin 
symmetric ($t_A$=0) and asymmetric (with asymmetry parameter, 
$t_A= 0.6$) nuclear 
%and hyperonic (with $f_s = 0.3$) 
matter.}
\label{Dens_kbi_ind_NM}
\end{figure}

\begin{figure}
\vskip -1.5in
\includegraphics[width=18cm,height=19cm]{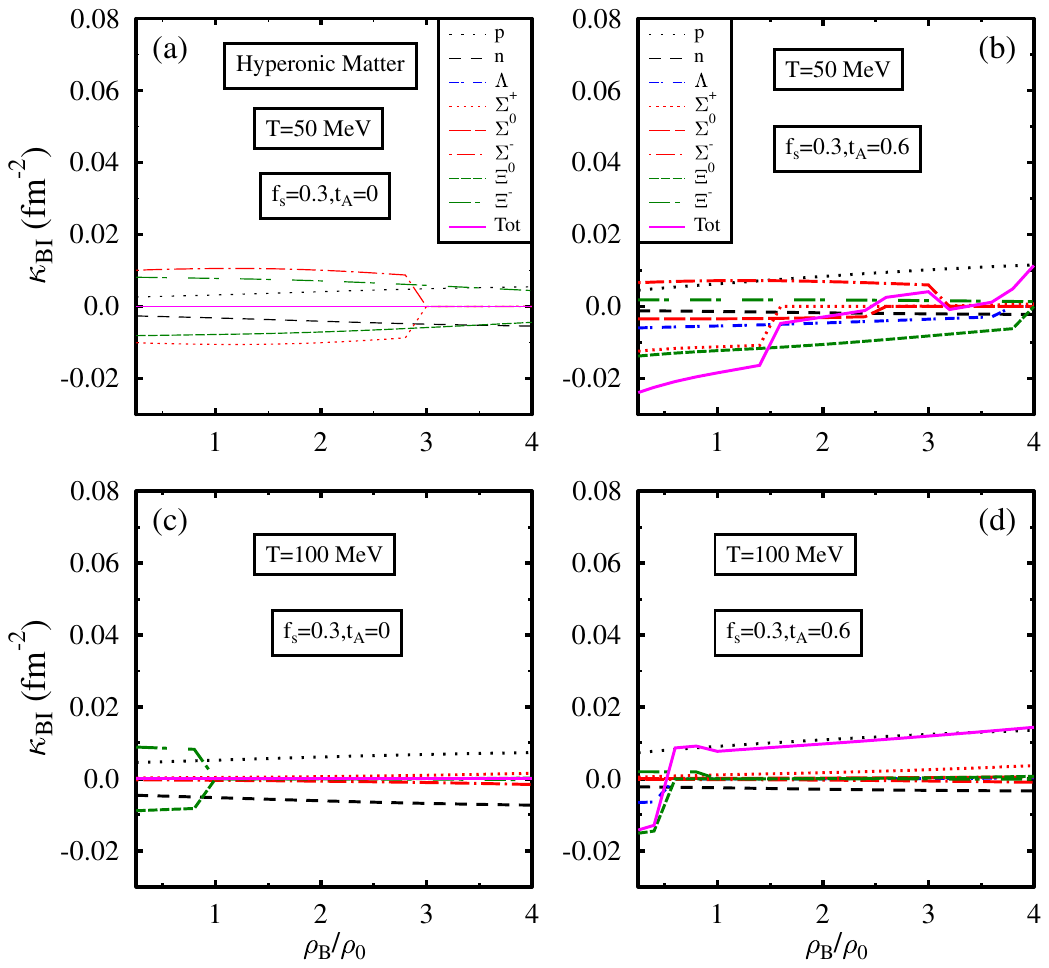} 
\vskip -0.5in
\caption{Diffusion coefficient, $\kappa_{BI}$ (in fm$^{-2}$),
is plotted as a function of the baryon density in units 
of nuclear matter saturation density,  $\rho_B/\rho_0$,
for different values of temperature for isospin 
symmetric ($t_A$=0) and asymmetric (with asymmetry parameter, 
$t_A = 0.6$) hyperonic (with $f_s = 0.3$) matter.}
\label{Dens_kbi_ind}
\end{figure}

\begin{figure}
\vskip -1.5in
\includegraphics[width=18cm,height=19cm]{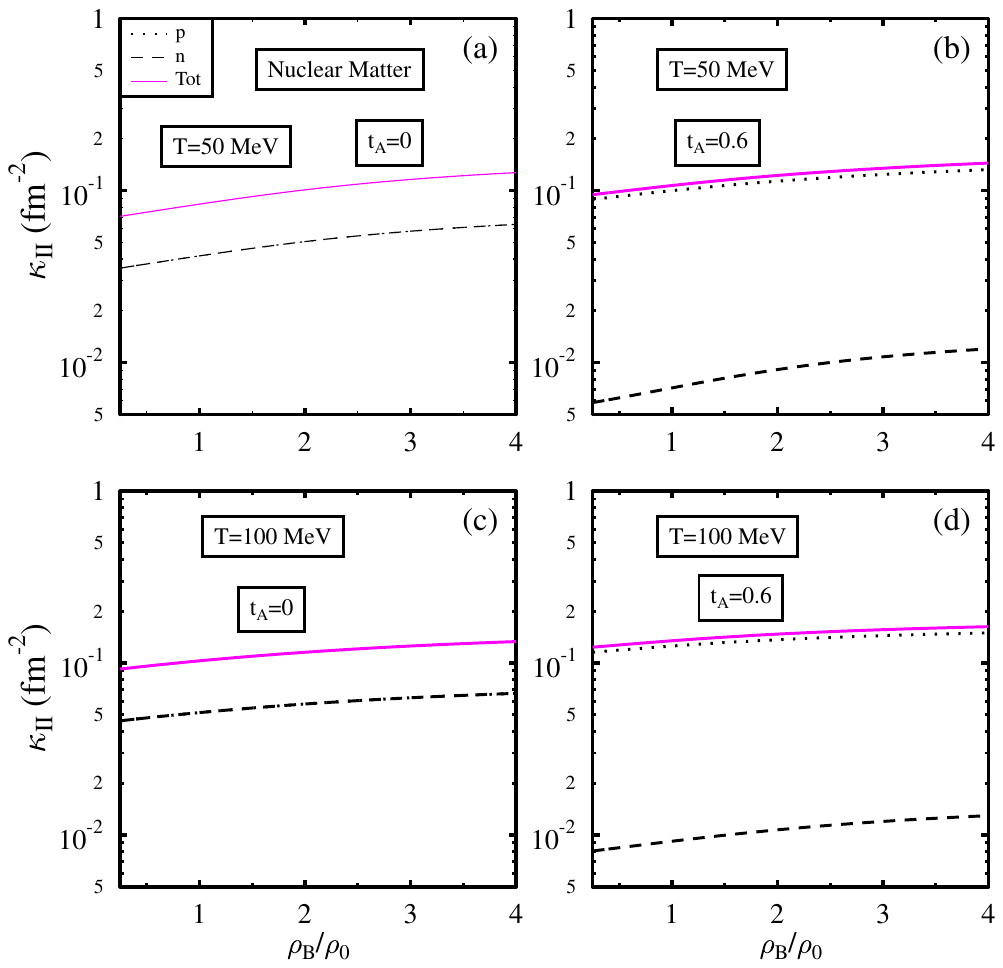} 
\vskip -0.5in
\caption{Diffusion coefficient, $\kappa_{II}$ (in fm$^{-2}$),
is plotted as a function of the baryon density in units 
of nuclear matter saturation density,  $\rho_B/\rho_0$,
for different values of temperature for isospin symmetric 
($t_A = 0$) and asymmetric (with asymmetry parameter, 
$t_A = 0.6$) nuclear 
%and hyperonic (with $f_s = 0.3$) 
matter.}
\label{Dens_kii_ind_NM}
\end{figure}

\begin{figure}
\vskip -1.5in
\includegraphics[width=18cm,height=19cm]{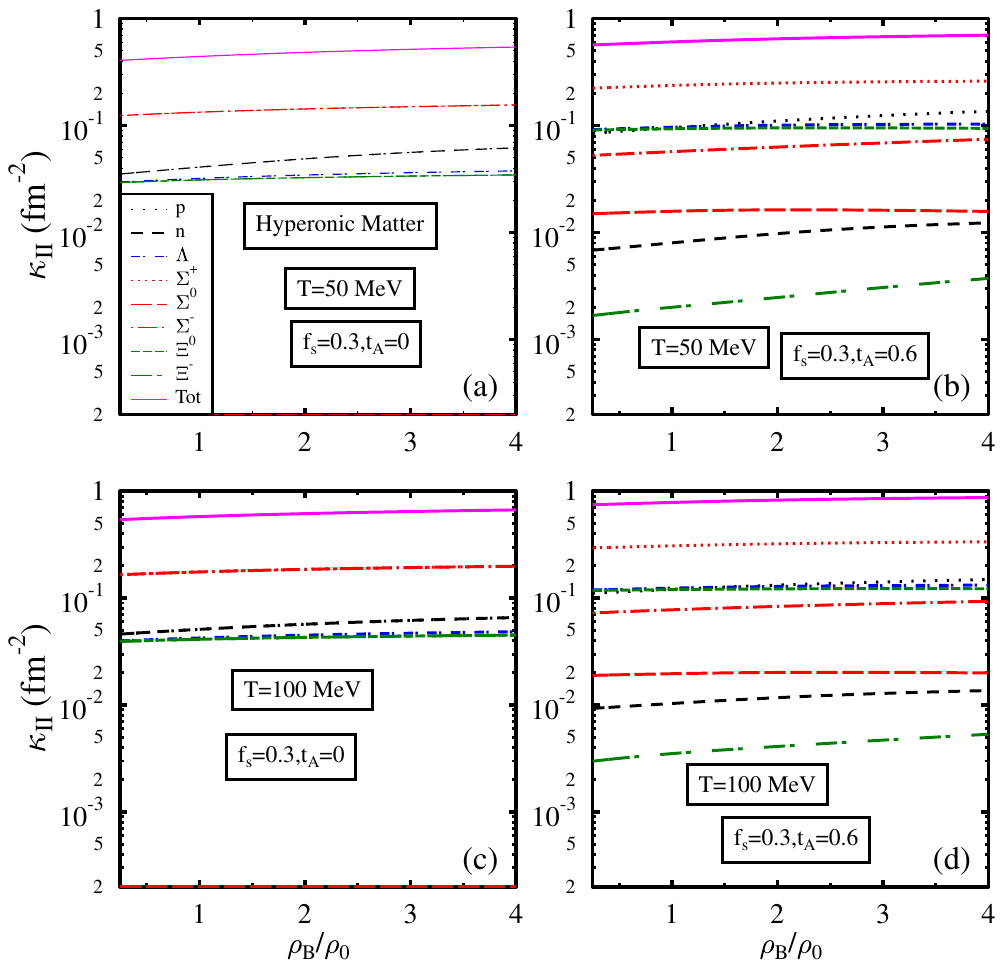} 
\vskip -0.5in
\caption{Diffusion coefficient, $\kappa_{II}$ (in fm$^{-2}$),
is plotted as a function of the baryon density in units 
of nuclear matter saturation density, $\rho_B/\rho_0$, 
for different values of temperature for isospin symmetric 
($t_A = 0$) and asymmetric (with asymmetry parameter, 
$t_A = 0.6$) hyperonic (with $f_s = 0.3$).} 
\label{Dens_kii_ind}
\end{figure}

\begin{figure}
\vskip -1.5in
\includegraphics[width=18cm,height=19cm]{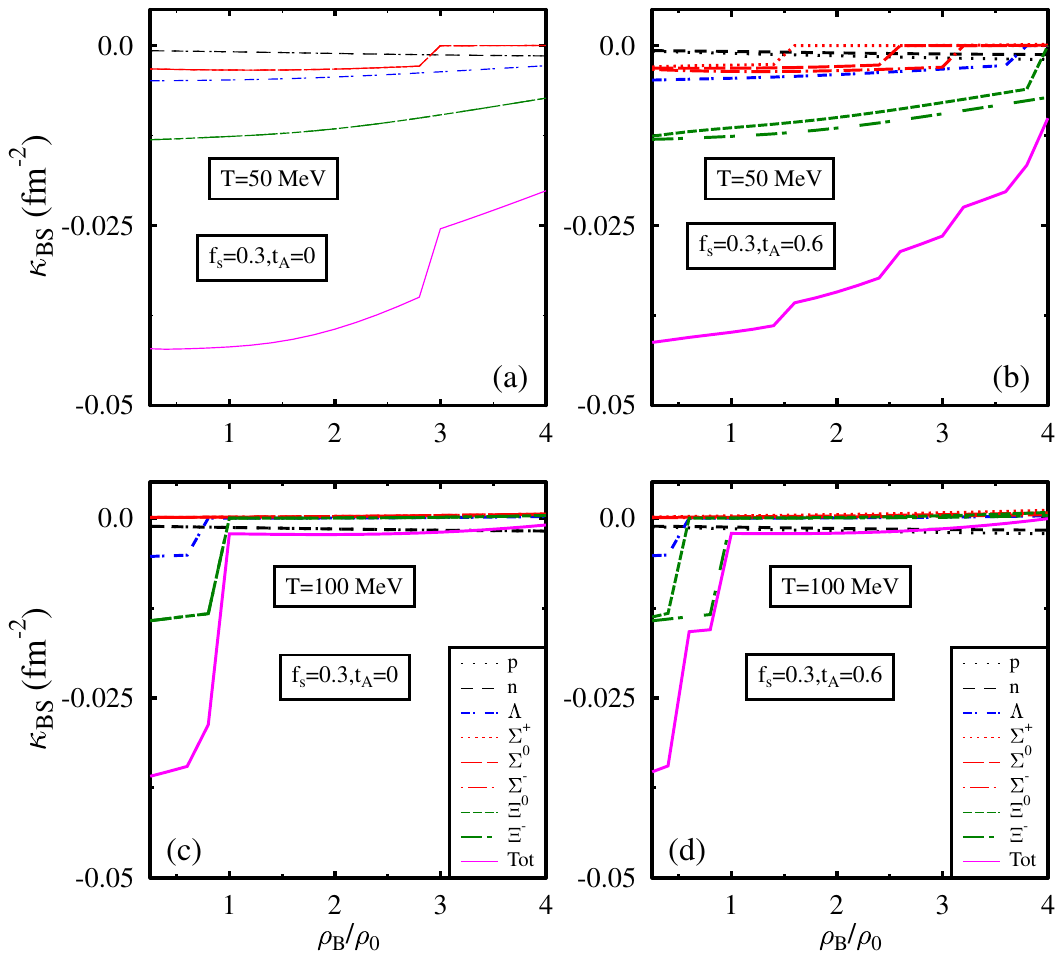} 
\vskip -0.5in
\caption{Diffusion coefficient, $\kappa_{BS}$ (in fm$^{-2}$),
is plotted as a function of  the baryon density in units 
of nuclear matter saturation density, $\rho_B/\rho_0$,
for different values of temperature for isospin symmetric 
($t_A = 0$) and asymmetric (with asymmetry parameter, 
$t_A = 0.6$) hyperonic matter (with $f_s = 0.3$).}
\label{Dens_kbs_ind}
\end{figure}

\begin{figure}
\vskip -1.5in
\includegraphics[width=18cm,height=19cm]{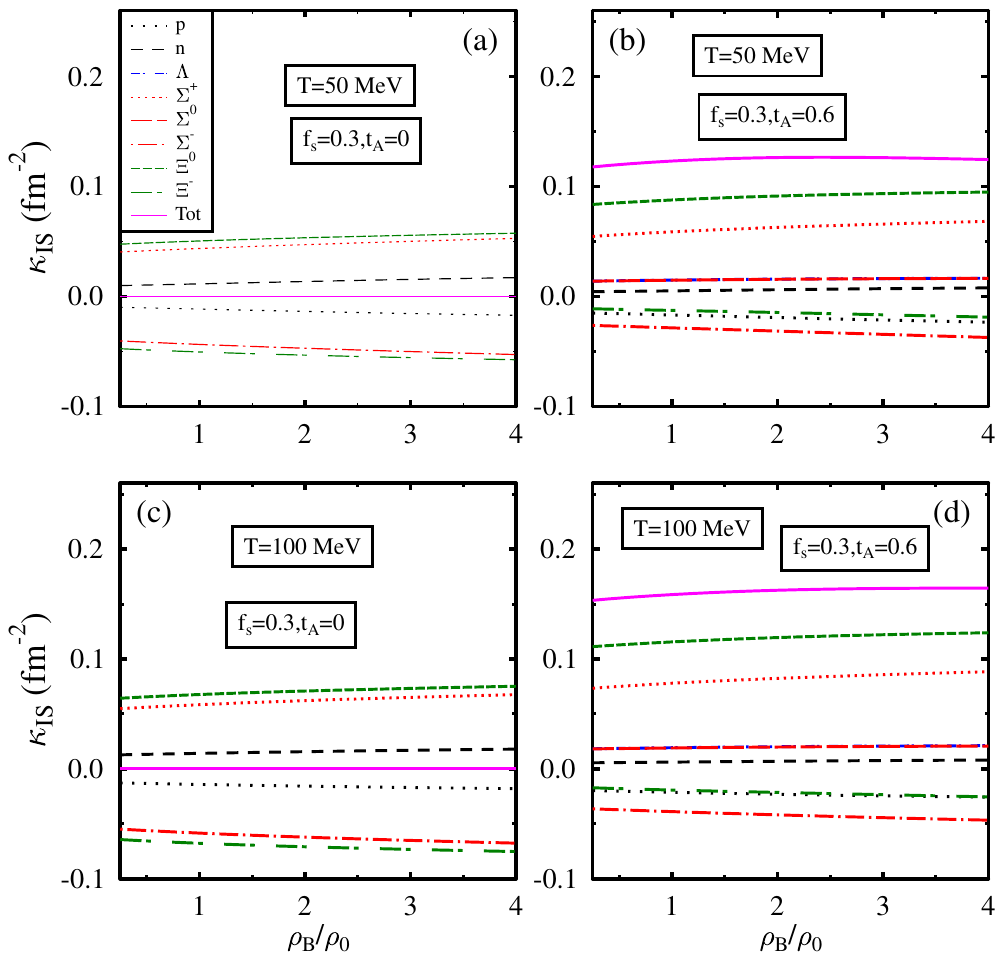} 
\vskip -0.5in
\caption{Diffusion coefficient, $\kappa_{IS}$ (in fm$^{-2}$),
is plotted as a function of the baryon density in units 
of nuclear matter saturation density, $\rho_B/\rho_0$,
for different values of temperature for isospin 
symmetric ($t_A$=0) and asymmetric (with asymmetry parameter, 
$t_A = 0.6$) hyperonic matter (with $f_s = 0.3$).}
\label{Dens_kis_ind}
\end{figure}

\begin{figure}
\vskip -1.5in
\includegraphics[width=18cm,height=19cm]{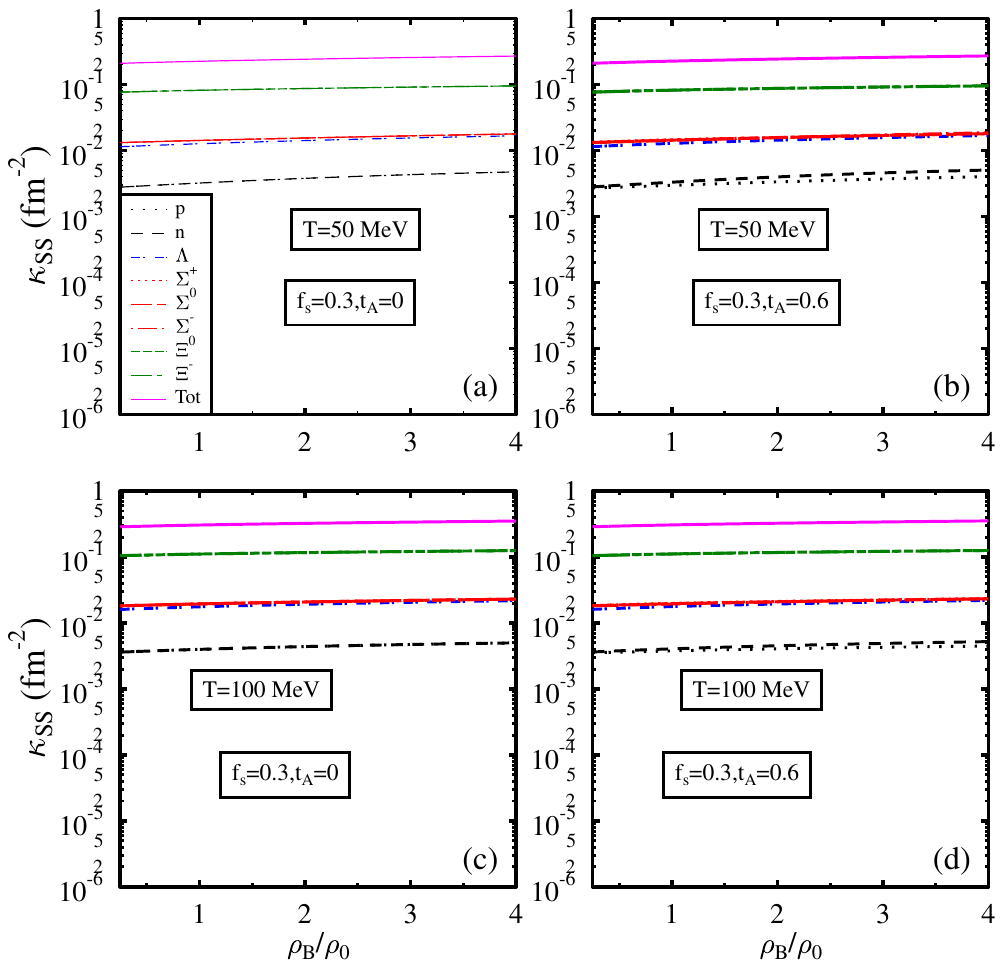} 
\vskip -0.5in
\caption{Diffusion coefficient, $\kappa_{SS}$ (in fm$^{-2}$),
is plotted as a function of  the baryon density in units 
of nuclear matter saturation density, $\rho_B/\rho_0$,
for different values of temperature for isospin symmetric 
($t_A = 0$) and asymmetric (with asymmetry parameter, 
$t_A = 0.6$) hyperonic matter (with $f_s = 0.3$).}
\label{Dens_kss_ind}
\end{figure}

\begin{figure}
\vskip -1.5in
\includegraphics[width=18cm,height=19cm]{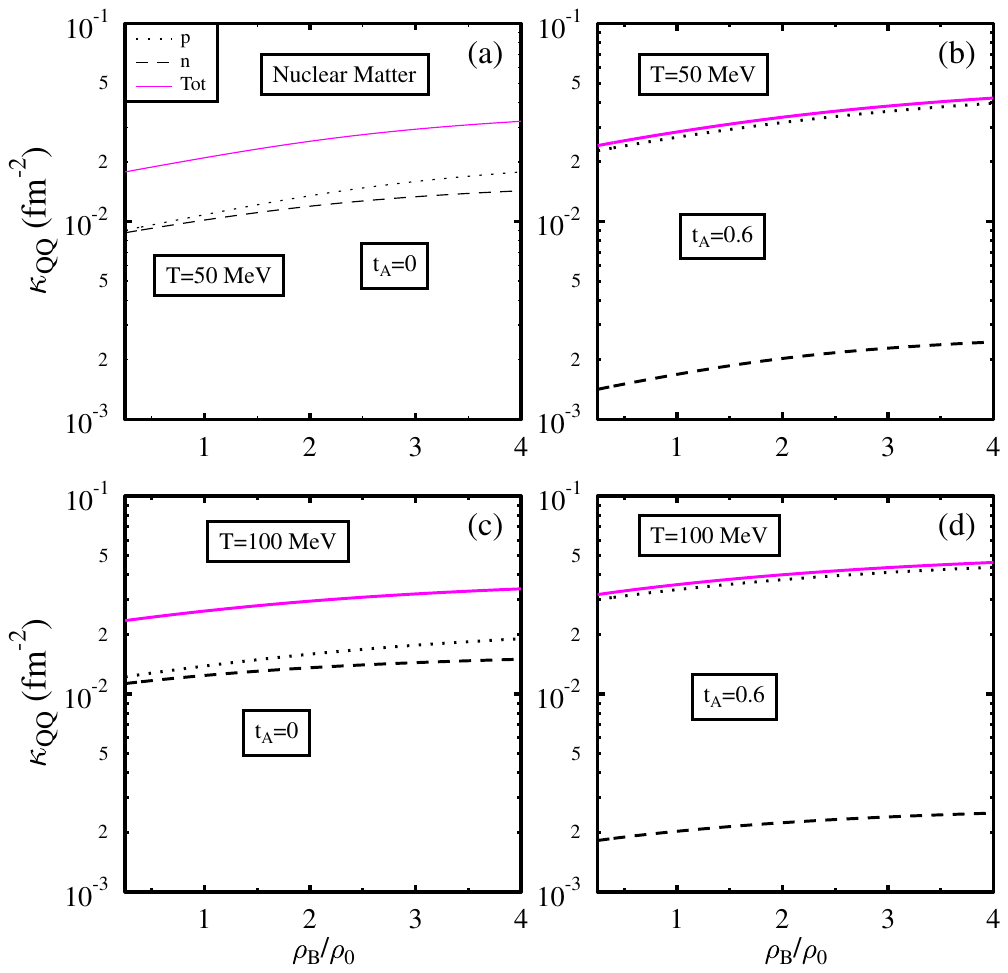} 
\vskip -0.5in
\caption{Diffusion coefficient, $\kappa_{QQ}$ (in fm$^{-2}$),
is plotted as a function of the baryon density in units 
of nuclear matter saturation density,  $\rho_B/\rho_0$,
for different values of temperature for isospin symmetric 
($t_A = 0$) and asymmetric (with asymmetry parameter, 
$t_A = 0.6$) nuclear 
matter.}
\label{Dens_kqq_ind_NM}
\end{figure}

\begin{figure}
\vskip -1.5in
\includegraphics[width=18cm,height=19cm]{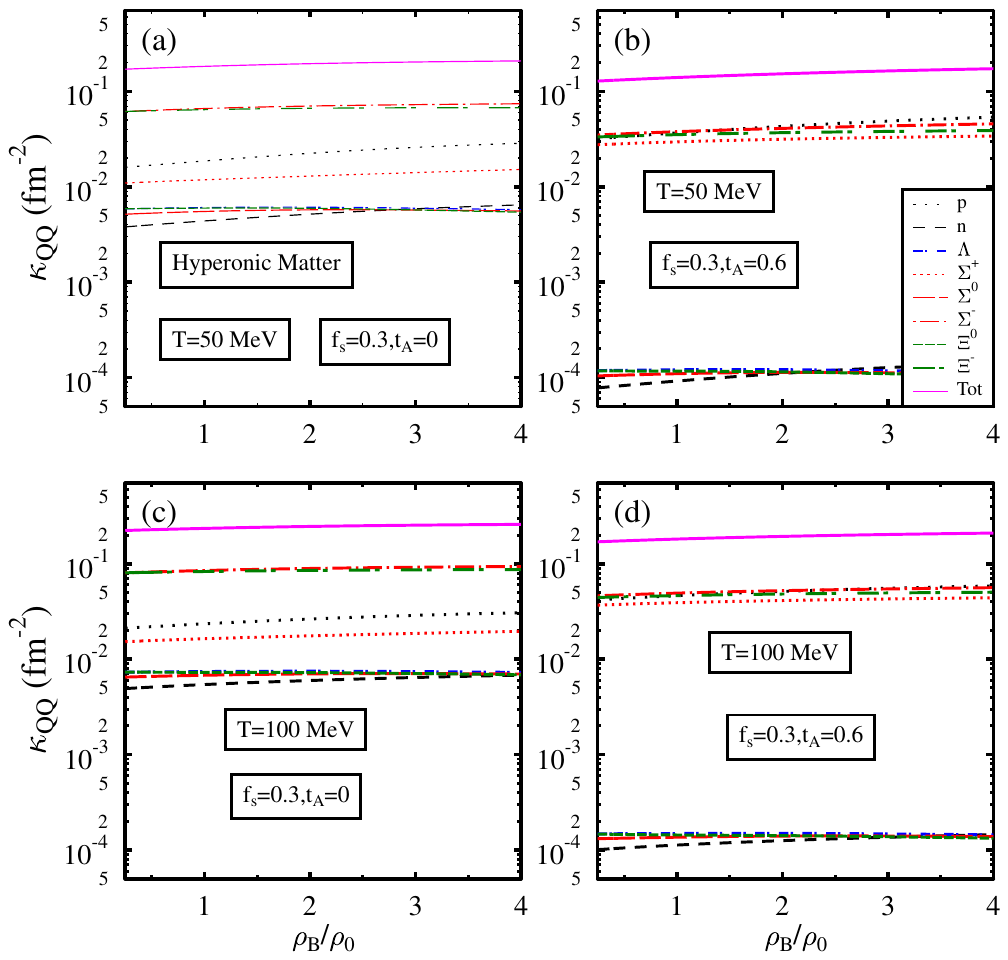} 
\vskip -0.5in
\caption{Diffusion coefficient, $\kappa_{QQ}$ (in fm$^{-2}$),
is plotted as a function of the baryon density in units 
of nuclear matter saturation density, $\rho_B/\rho_0$, 
for different values of temperature for isospin symmetric 
($t_A = 0$) and asymmetric (with asymmetry parameter, 
$t_A = 0.6$) hyperonic (with $f_s = 0.3$).} 
\label{Dens_kqq_ind_SHM}
\end{figure}

\begin{figure}
\vskip -2.in
\includegraphics[width=18cm,height=19cm]{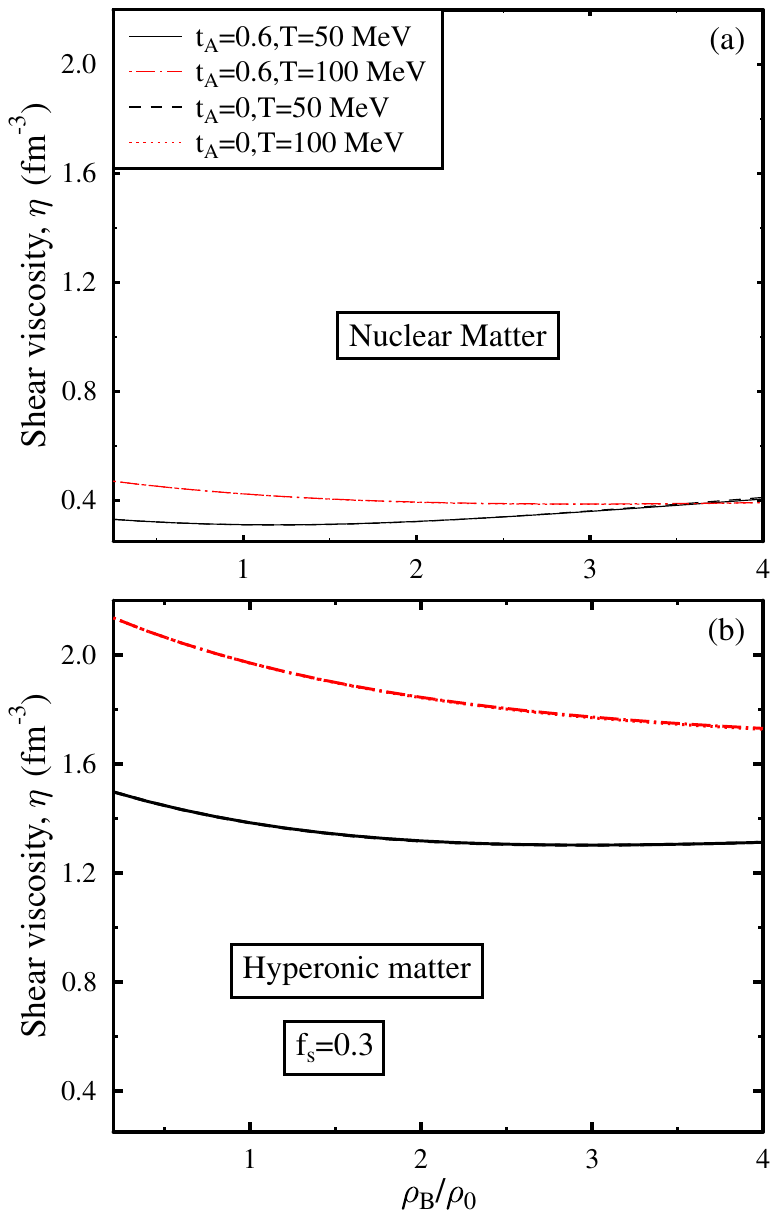} 
\vskip -0.1in
\caption{Shear viscosity, $\eta$ (in fm$^{-3}$),
is plotted as a function of the baryon density in units 
of nuclear matter saturation density,  $\rho_B/\rho_0$,
for different values of temperature for isospin symmetric 
($t_A = 0$) and asymmetric matter (with asymmetry parameter, 
$t_A = 0.6$) in nuclear matter and hyperonic matter (with $f_s = 0.3$).}
\label{Dens_shear_visc_NM_SHM}
\end{figure}

\begin{figure}
\vskip -2.in
\includegraphics[width=18cm,height=19cm]{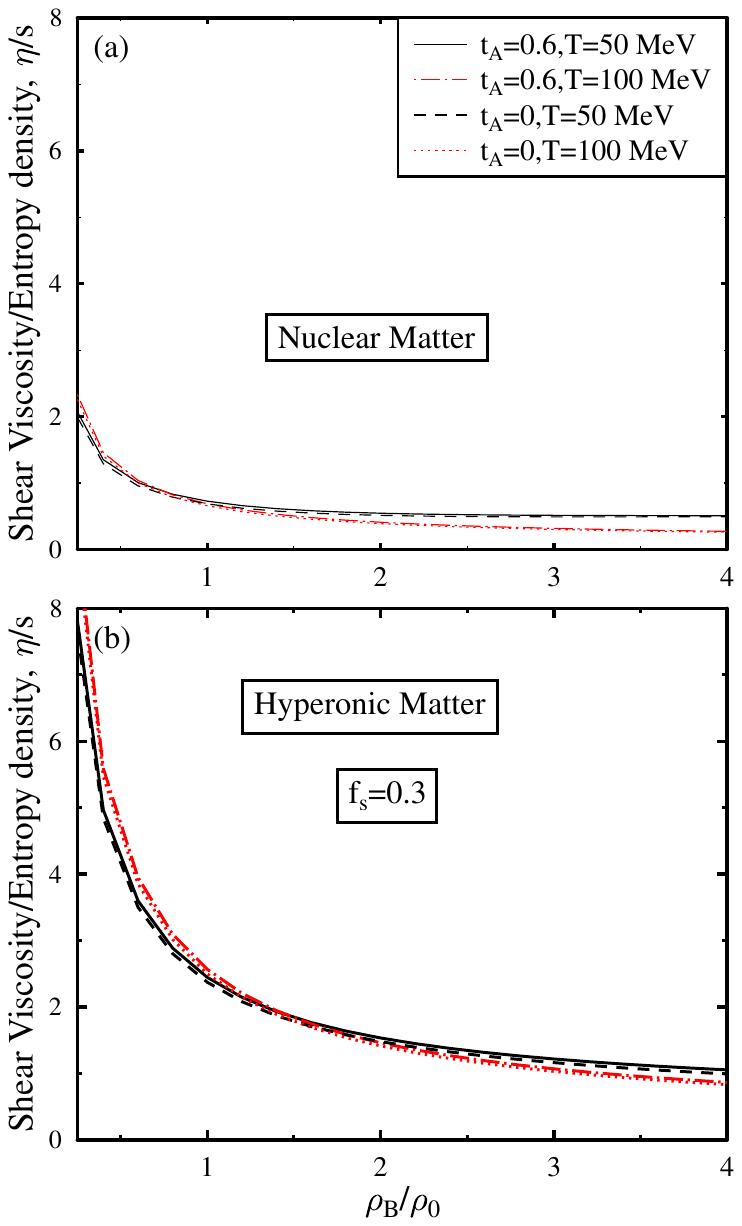} 
\vskip -0.1in
\caption{Shear viscosity per entropy density, $\eta/s$,
is plotted as a function of the baryon density in units 
of nuclear matter saturation density, $\rho_B/\rho_0$,
for different values of temperature, for isospin symmetric 
($t_A = 0$) and asymmetric matter (with asymmetry parameter, 
$t_A = 0.6$) in nuclear matter and hyperonic matter (with $f_s = 0.3$).}
\label{Dens_shear_visc_entr_NM_SHM}
\end{figure}

\section{Results and discussions}
\label{sec:results-discussion}

We discuss the results of the present investigation
of the thermodynamic and transport properties 
of hot isospin asymmetric strange hadronic matter.
In Fig.~\ref{Dens_EOS_SHM}, the pressure (in MeV/fm$^{3}$) 
is plotted as a function of the energy density (in MeV/fm$^{3}$) 
at given values of the temperature for the
isospin asymmetric (with $t_A=0.6$) nuclear matter as well as for strange hadronic matter 
(with strangeness fraction, $f_s = 0.3$) and compared with 
the isospin symmetric matter.
The isospin asymmetry leads to higher pressure as compared to the symmetric matter.
The effect of strangeness leads to a softer equation of state, similar to the observation for the neutron star matter in the presence of hyperons
\cite{Glendenning_Schaffner_PRC60_025803_1999}.

Figure \ref{Dens_Entr_SHM} shows the baryon density dependence of the entropy 
density (in fm$^{-3}$) for nuclear and hyperonic matter.
There is an observed increase in the entropy density
in the presence of strangeness in the medium due to degrees of freedom from the hyperons
added to the entropy density.
The effect of isospin asymmetry is observed to lead to a lower value
compared to symmetric matter for both
nuclear matter and hyperonic matter. However, the effect
of isospin asymmetry is minor.

In Fig.~\ref{Dens_meffa}, the effective masses of the 
nucleons and the hyperons (given by Eq.~(\ref{ameff}))
in strange hadronic matter (with strangeness fraction,
$f_s = 0.3$) are plotted as functions of the baryon density in 
units of nuclear matter saturation density.
For values of the temperature, $T=50$ and $100$ MeV,
the masses are shown in panels (a) and (c)
for symmetric ($t_{A} = 0$), and in panels (b) and (d)
for asymmetric (with $t_A = 0.6$) matter.
In the symmetric matter, the baryon masses are degenerate
within a given isospin multiplet (($p$, $n$), $(\Sigma^-,\Sigma^0,\Sigma^+)$,
$(\Xi^-,\Xi^0)$), which become nondegenerate
in the presence of isospin asymmetry in the medium
due to the interaction with the isovector scalar 
($\delta$) meson. Masses within the isospin multiplet are higher for the
higher values of the isospin projection operator.
The nucleons are observed to have a much larger drop 
in the medium as compared to hyperons.

In Fig.~\ref{Dens_rhoa}, for different values of temperature,
we plot the number densities (in fm$^{-3}$)
of the nucleons and hyperons $\rho_a$ with 
$a=p$, $n$, $\Lambda$, $\Sigma^{\pm,0}$, and $\Xi^{-,0}$, in strange hadronic matter
(with strangeness fraction, $f_s = 0.3$) 
as functions of the baryon density
in units of nuclear matter saturation density. For isospin symmetric 
($t_A = 0$) matter, the number densities of the baryons
within an isospin multiplet are the same. However, 
in the presence of isospin asymmetry in the medium,
the number densities of the baryons within a given  
multiplet are no longer the same.
The density dependence of the relaxation 
time associated with diffusion of a given species of baryon $\tau_a^d$
given by Eq.~(\ref{v_almb_a_d}) is plotted 
both for the symmetric and asymmetric matter
in Figs.~\ref{Dens_tau_NM} and \ref{Dens_tau_Nuc_Hyp}
for nuclear matter and hyperonic matter, respectively.
The density trend of the relaxation time is predominantly 
determined by its mean free path ($\lambda_a$), 
which is inversely proportional to its number density, 
$\rho_a$. 
In asymmetric nuclear matter, due to a higher value of $\rho_n$
as compared to $\rho_p$, the relaxation time for neutron
is observed to be smaller than for protons.

In Fig.~\ref{Dens_tau_Nuc_Hyp}, the relaxation times of the nucleons
and hyperons associated with diffusion  $\tau_a^d$  given by Eq.~(\ref{v_almb_a_d})
are shown for the strangeness fraction
of $f_s = 0.3$. The density dependence remains the same
as for the case of nuclear matter. However, the relaxation
times for the hyperons are much larger due to the smaller
values of their number densities. 
For asymmetric hyperonic matter at $T = 50$ MeV (see panel (b) in Fig.~\ref{Dens_rhoa}), the relaxation time of $\Xi^0$ exhibits a density dependence that differs from the usual trend: rather than monotonically decreasing with increasing baryon density $\rho_B$, it first drops with density and then increases as the density is further raised.
This is a reflection of its density showing
the opposite trend (see panel (b) of Fig.~\ref{Dens_rhoa}) 
and, as mentioned before,
the density behavior of $\tau_a^d
(=\lambda_a^d/\langle v_a^d\rangle)$ 
is predominantly due to the mean free path, which is inversely
proportional to its number density. It might be noted here that 
the relaxation time corresponding to the shear viscosity 
given by Eq.~\eqref{v_almb_a_shear} is extremely close to 
the value of the relaxation time corresponding to diffusion $\tau_a^d$ given by Eq.~\eqref{v_almb_a_d}. This reflects the fact that the contribution from the anti-particle distribution function, which is zero for zero temperature, is negligible for the temperatures of $50$ and $100$ MeV considered in the present work.

In the following, we shall show the density dependence
of the diffusion coefficients $\kappa_{ij}$
in hot strange hadronic matter,
arising due to the multiple conserved charges, $q_i$, $i=1$, $2$, and $3$  
corresponding to the baryon ($B$), isospin ($I$),
and strangeness ($S$) charges. We might note here, 
in the expression for the diffusion coefficients
as given by Eq.~(\ref{kappa_ij_vd}),
the integrand depends on the product of the
bracketed quantities
${\cal B}_{ia}\equiv \left(q_{ia}-\frac{\rho_{qi}}{\cal W}E_{a}\right)$ 
and ${\cal B}_{ja}\equiv \left(q_{ja}-\frac{\rho_{qj}}{\cal W}E_{a}\right)$, 
which need not be positive definite for the case of $q_i \ne q_j$. 
Assuming the momenta to be small as compared to the effective mass
of $a$-th baryon, the expression $E_a$ in the integrand
can be approximated as $E_a=E_a^*\pm 
(g_{\omega a}\omega +g_{\rho a}\rho +g_{\phi a}\phi)
\sim m_a^*\pm (g_{\omega a}\omega +g_{\rho a}\rho
+g_{\phi a}\phi)$ in the leading order.
The density behavior of the contribution of the $a$-th baryon to
the diffusion coefficients can be understood from the 
the density dependences of the product ${\cal B}_{ia} \cdot {\cal B}_{ja}$ 
(using the approximate expression of $E_a$),
and of the relaxation time $\tau_a^d$.
It is observed that the diffusion coefficients, 
$\kappa_{ij}$, which are the sum of contributions of all the baryons
with different effective masses, chemical potentials,
and relaxation times can have quite nontrivial density
dependence.

In Fig.~\ref{Dens_kbb_ind_NM}, the diffusion coefficient 
$\kappa_{BB}$ (in units of ${\rm {fm}}^{-2}$) is plotted 
as the function of the baryon density in units 
of nuclear matter saturation density $\rho_B/\rho_0$ for given temperatures
for both symmetric and asymmetric nuclear matter
along with individual contributions from the nucleons.
In asymmetric nuclear matter, the contributions from
protons are observed to be larger as compared
to the neutrons, due to the larger values of the
relaxation time. For lower values of the temperature,
$T = 50$ and $100$ MeV, an increase
in the diffusion coefficient with density is observed.
In the presence of the hyperons,
as can be seen in Fig. \ref{Dens_kbb_ind_SHM}, 
the density trend is similar to that of the nuclear matter.
However, the value of $\kappa_{BB}$
is enhanced substantially due to contributions
from hyperons in addition to the nucleons.

In Figs.~\ref{Dens_kbi_ind_NM} and \ref{Dens_kbi_ind}, 
$\kappa_{BI}$ is plotted as a function of density 
for nuclear and hyperonic matter,
both for symmetric and asymmetric matter.
In symmetric nuclear matter, as the isospin charge density 
vanishes,  the contributions are equal and opposite
for protons and neutrons, yielding the value
of $\kappa_{BI}$ to be zero for symmetric nuclear matter,
as can be seen from (a) and (c) of Fig.~\ref{Dens_kbi_ind_NM} for $T=50$ and $100$ MeV, respectively. In asymmetric nuclear matter, however,
as can be seen from (b) and (d) of the same figure,
there is a larger contribution from 
protons compared to neutrons, which has predominantly
higher relaxation time relative to the neutron.
In the presence of the asymmetry, 
in nuclear matter, the contributions from the neutrons
(protons) still remain negative (positive), but
the total $\kappa_{BI}$ is observed to be positive,
can be seen in (b) and (d) of Fig.~\ref{Dens_kbi_ind_NM}.
For $T=50$ and $100$ MeV,
there is an increase with density.
In symmetric hyperonic matter, the contributions
from the isospin multiplets add up to give
the total contribution to be zero for $\kappa_{BI}$
as can be seen from (a) and (c) of Fig. ~\ref{Dens_kbi_ind}. 
However, in asymmetric hyperonic matter,
with the individual contributions as positive or negative,
the total value is observed to be  negative at low densities
for $T = 50$ and $100$ MeV, whereas there is a change of sign
at higher values of the densities, as can be seen from 
(b) and (d) of Fig. ~\ref{Dens_kbi_ind}.

Figure \ref{Dens_kii_ind_NM} shows the density dependence of $\kappa_{II}$
in nuclear matter, along with individual contributions
from the nucleons. The trend of increasing with 
increase in density is similar to that observed for
$\kappa_{BB}$ (shown in Fig.~\ref{Dens_kbb_ind_NM}). 
 In asymmetric nuclear matter, $\kappa_{II}$ has
a much larger value for protons, which has a much larger
relaxation time as compared to the neutron.
For the hyperonic matter, the value of $\kappa_{II}$
is appreciably larger as compared to nuclear matter
due to additional contributions from the hyperons,
as can be seen from Fig.~\ref{Dens_kii_ind}.

For strange hadronic matter, the diffusion coefficients
$\kappa_{BS}$, $\kappa_{IS}$, and $\kappa_{SS}$ are
plotted as functions of density in
Figs.~\ref{Dens_kbs_ind}, \ref{Dens_kis_ind},
and \ref{Dens_kss_ind}, respectively. The values 
of $\kappa_{BS}$ are observed to be negative 
for the lower temperatures, $T = 50$ and $100$ MeV
though the density dependence is similar for the symmetric 
and asymmetric cases. As can be seen from the expression
of $\kappa_{IS}$, the total of the contributions
within the isospin multiplets add to zero for the isospin
symmetric hyperonic matter. However, for isospin asymmetric
matter (with asymmetric parameter $t_{A} = 0.6$), the values
are observed to be positive at all densities 
for the considered temperatures. The values
of $\kappa_{SS}$ are observed to be positive and
are larger for larger values of the temperature.

The results for the diffusion coefficients
as shown above correspond to the elements of the diffusion matrix with the baryon number, isospin, and strangeness charges.
It is worthwhile to investigate the density dependence
of the diffusion coefficient, $\kappa_{QQ}$ ($q_i=q_j=Q$) 
corresponding to the conserved electric charge, which is related to the electrical conductivity of the hadronic matter. The diffusion coefficients $\kappa_{QQ}$ are obtained 
from Eq. ({\ref{kappa_ij_vd}}), using the electric charge 
density,
$\rho_Q\left(=\sum_a Q_a \rho_a \right)$, where
$Q_a$ is the electric charge of the 
$a$-th baryon. The diffusion coefficient, $\kappa_{QQ}$, 
is plotted in Figs. 
\ref{Dens_kqq_ind_NM} and \ref{Dens_kqq_ind_SHM}
for the nuclear matter and hyperonic matter, respectively. For symmetric nuclear matter, we get a smaller value of $\kappa_{QQ}$ than $\kappa_{II}$ for protons, due to the non-zero charge density $\rho_Q$. For the case of asymmetric nuclear matter, due to the larger value of the relaxation time for protons compared to neutrons, we observe a much larger value of $\kappa_{QQ}$, as can be seen from Fig. \ref{Dens_kqq_ind_NM}. This is similar to the case of $\kappa_{II}$ shown in Fig. \ref{Dens_kii_ind_NM}. In the presence of the isospin asymmetry, for hyperonic matter, similar to the case of nuclear matter, one still observes that the contribution from the $a$-th baryon to $\kappa_{QQ}$ is predominantly due to the behavior of the relaxation time $\tau_{a}^d$, which is inversely proportional to its density. The inclusion of hyperons is observed to lead to an enhancement in the value of $\kappa_{QQ}$ compared to nuclear matter. 

The shear viscosity coefficient $\eta$ is presented in Fig.~\ref{Dens_shear_visc_NM_SHM}.
For hyperonic matter, $\eta$ is significantly larger than for nuclear matter, owing to the additional hyperonic degrees of freedom.
Isospin-asymmetry effects are marginal as observed in nuclear matter.
The coefficient of shear viscosity to entropy-density ratio $\eta/s$ is shown in Fig.~\ref{Dens_shear_visc_entr_NM_SHM};
it exhibits a similar density dependence with or without hyperons in the system, however, the values are larger for hyperonic matter at high densities.

\section{Summary}
\label{sec:summary}

In this paper, the thermodynamics and transport properties
of hot isospin asymmetric strange baryonic matter
have been studied within a chiral SU(3) model.
Strangeness softens the equation of state, whereas the entropy density is enhanced when hyperons are included
in the nuclear medium. The diffusion coefficients
associated with multiple conserved quantities, namely
baryon, isospin, and strangeness charges, as well as
the effects of hyperons on the coefficient
of the shear viscosity are studied. 
The inclusion of the hyperons is observed to lead to much larger values of the 
diffusion coefficients, $\kappa_{BB}$ and $\kappa_{II}$ as compared
to nuclear matter. 
The contributions of the baryons 
to the non-diagonal diffusion matrix elements ($\kappa_{BI}$, 
$\kappa_{BS}$ and $\kappa_{IS}$), which take either positive
or negative values,
are observed to lead to quite non-trivial density dependence
for these coefficients. On the other hand, the diagonal diffusion
coefficients, $\kappa_{BB}$, $\kappa_{II}$ and $\kappa_{SS}$,
are always positive definite. We have also investigated the electric charge diffusion coefficient $\kappa_{QQ}$ in the hot asymmetric hadronic medium. The inclusion of hyperons is observed to lead to an enhancement in the value of $\kappa_{QQ}$ as compared to nuclear matter. It is still observed that the density behavior of the individual baryons is predominantly due to the mean free path in the expression of the relaxation time, which is inversely proportional to their number densities. The effects 
on the diffusion coefficients from the isospin asymmetry as well as 
strangeness in the medium are observed to be quite appreciable.
The shear viscosity coefficient has a large enhancement in the presence of strangeness due to 
additional contributions from the hyperons; however, the effects 
due to isospin asymmetry remain marginal both in nuclear and hyperonic 
matter. The present study can be relevant
for hydrodynamic analyses of experimental observables
in relativistic heavy-ion 
collisions in the compressed baryonic matter (CBM) experiment 
at the FAIR facility at GSI as well as in the future J-PARC-HI program in Japan.

\acknowledgements
One of the authors (AM) would like to acknowledge the kind hospitality
at the Department of Engineering and Applied Science,
Faculty of Science and Technology, Sophia University, Japan,
where the work was initiated.

\end{document}